%% file: IONconf_template.tex
\documentclass[letterpaper,times]{IONconf}
\usepackage[hidelinks]{hyperref}

\usepackage[noadjust]{cite}
\usepackage{amsmath,amssymb,amsfonts}
\usepackage{glossaries}
\usepackage{multicol}

\usepackage{graphicx}
\usepackage{textcomp}
\usepackage{xcolor}
\usepackage[capitalise]{cleveref}
\usepackage{siunitx}
\usepackage{subcaption}
\usepackage{tabularx}
\usepackage{colortbl}
\usepackage{algpseudocode}
\usepackage{algorithm}
\usepackage{xurl}
\usepackage{flushend}
\usepackage{tablefootnote}
\usepackage{bm}
\usepackage[raggedrightboxes]{ragged2e}

\DeclareSIUnit[]\gforce{\mathcal{g}}

\crefname{subsection}{Subsection}{Subsections}

\DeclareSIUnit\operations{op}
\DeclareSIUnit\gforce{g}

\usepackage[colorinlistoftodos,prependcaption,textsize=tiny]{todonotes}
\input{acronyms.tex}

\usepackage{url}

\usepackage{natbib}

\title{Consumer INS Coupled with Carrier Phase Measurements for GNSS Spoofing Detection}

\author{
    Tore Johansson, Marco Spanghero, Panos Papadimitratos \\ \textit{Networked Systems Security (NSS) Group -- KTH Royal Institute of Technology, Stockholm, Sweden}
    \vspace{1mm} \\%
    torej@kth.se,marcosp@kth.se,papadim@kth.se
    }

\begin{document}

\maketitle

\section*{biography}


\biography{Tore Johannson}{is an Embedded software developer in the defense industry. He received a M.Sc. in Embedded Systems at KTH, Royal Institute of Technology, Sweden. His area of interest include navigation systems, control systems and system-level security.}

\biography{Marco Spanghero}{B.Sc. in Electronics Engineering from Politecnico di Milano, M.Sc. in Embedded Systems at KTH. He is currently a Ph.D. candidate with the Networked Systems Security (NSS) group at KTH Royal Institute of Technology, Stockholm, Sweden in satellite-based robust and reliable navigation and timing.}

\biography{Panos Papadimitratos}{is a professor with the School of Electrical Engineering and Computer Science (EECS) at KTH Royal Institute of Technology, Stockholm, Sweden, where he leads the Networked Systems Security (NSS) group. He earned his Ph.D. degree from Cornell University, Ithaca, New York, in 2005. His research agenda includes a gamut of security and privacy problems, with an emphasis on wireless networks. He is an IEEE Fellow, an ACM Distinguished Member, and a Fellow of the Young Academy of Europe.}

\section*{Abstract}

Global Navigation Satellite Systems enable precise localization and timing even for highly mobile devices, but legacy implementations provide only limited support for the new generation of security-enhanced signals. Inertial Measurement Units have proved successful in augmenting the accuracy and robustness of the GNSS-provided navigation solution, but effective navigation based on inertial techniques in denied contexts requires high-end sensors. However, commercially available mobile devices usually embed a much lower-grade inertial system. To counteract an attacker transmitting all the adversarial signals from a single antenna, we exploit carrier phase-based observations coupled with a low-end inertial sensor to identify spoofing and meaconing. By short-time integration with an inertial platform, which tracks the displacement of the GNSS antenna, the high-frequency movement at the receiver is correlated with the variation in the carrier phase. In this way, we identify legitimate transmitters, based on their geometrical diversity with respect to the antenna system movement. We introduce a platform designed to effectively compare different tiers of commercial INS platforms with a GNSS receiver. By characterizing different inertial sensors, we show that simple MEMS INS perform as well as high-end industrial-grade sensors. Sensors traditionally considered unsuited for navigation purposes offer great performance at the short integration times used to evaluate the carrier phase information consistency against the high-frequency movement. Results from laboratory evaluation and through field tests at Jammertest 2024 show that the detector is up to 90\% accurate in correctly identifying spoofing (or the lack of it), without any modification to the receiver structure, and with mass-production grade INS typical for mobile phones. 

\section{Introduction}
\label{sec:introduction}





\gls{gnss} constellations are the most common providers of precise location and time for a wide gamut of devices. Signals designated for civilian usage mostly lack security features to stop adversarial manipulation and interference, with the exception of \gls{osnma} in the Galileo system. The lower entry cost for effective adversaries and the availability of budget, high-performance \gls{sdr} jointly with open-source tools for signal falsification made this threat model significant even in civilian receivers, both in the case of spoofing \cite{HumphreysAssessingSpoofer,Humphreys2012,HuangL2015} and meaconing \cite{LenhartSP:C:2022,10.1145/3558482.3590186}. In this context, several spoofing cases (intentional or unintentional) have been documented that caused misbehavior in navigation systems \cite{7471534,SkytruthJamming,spirentSpoofing}. 

To strengthen the current civilian navigation infrastructure, Galileo \gls{osnma} (\cite{Gotzelmannnavi.572,8714151,Cucchi2021AssessingReceiver}) and GPS Chimera (\cite{WOS:000419292302031,WOS:000874785704023}) modify the structure of the signal in space adding authenticated navigation information and, possibly, authenticated spreading codes. While the stronger approach relying on authenticated spreading codes will significantly raise the bar for any unsophisticated spoofing attack, the authentication of the navigation frames only partially addresses the spoofing issue, leaving the receiver vulnerable to signal replay and relay \cite{LenhartSP:C:2022,ZhangLP:J:2022}. Adoption of navigation message authentication, whose security hardening does not require modifications to the physical layer signal structure, is accelerating toward the public service phase of \gls{osnma} (\cite{ubx_osnma,10139953}) but devices already deployed are not guaranteed to be upgradable.

Approaches that combine measurements from the GNSS receiver with an \gls{imu} \cite{Curran2017OnTU}, focus on the consistency of the device movement between the \gls{gnss} solution and the inertial system estimation.
In combination with more advanced measurements provided by commercial GNSS modules (i.e., raw observations of pseudoranges, code, and carrier phase), fusion of multiple sources of information is possible even in low \gls{swap} mobile devices \cite{Lee2022GNSSFD,Sharma2021TimeSynchronizedGD}. Nevertheless, general purpose \gls{imu}s typically available in mobile platforms are unsuitable for navigation due to their large intrinsic errors, when operating in a truly denied context.  

When applied to moving targets, spoofing requires a higher level of sophistication to be successful, in particular for highly synchronized and smooth takeover attacks \cite{HumphreysAssessingSpoofer}. Practically, it is generally unfeasible for an adversary to accurately determine the carrier phase of individual signals if the victim is moving rapidly. This would require real-time knowledge of the victim antenna phase center position with cm-level accuracy. This makes the adversary unable to perfectly match the carrier variations due to high-frequency receiver antenna motion, which, however, can be accurately measured by the victim relying on short-time inertial methods. 

This motivates our investigation here on how a mobile platform can leverage low-cost \gls{imu} and raw \gls{gnss} measurements to efficiently validate the point of origin of the satellite signals with a single antenna, relying on its high-frequency movement. We show how carrier phase structure estimation with a short-term inertial determination of the antenna movement enables distinguishing spoofed from real signals, with very limited assumption on the type of movement. 
Commercial mass-market receivers support multi-Hz update rates but generally are limited to \SI{25}{\hertz}. More advanced receivers reach higher measurement rates (generally limited to \SI{100}{\hertz}), but the processing power required to run the algorithm would not allow real-time operation at such a high sampling rate. 

Our detection method can run as soon as satellites are available, as it is decoupled from the availability of a \gls{pnt} solution. Additionally, it is completely agnostic to the receiver's position and state, only requiring that carrier phase measurements are available. In other words, our method validates signals that have not yet been used by the GNSS receiver in the \gls{pnt} solution, in contrast to traditional \gls{raim} methods.
 
Specifically, our contributions are:
\begin{itemize}
    \item Improved carrier phase-based spoofer detection, relying on high-frequency antenna movement with generic mechanization
    \item Real-time tracking of arbitrary movements of a GNSS antenna, practical even for low-cost \gls{imu} sensors
    \item A novel platform to evaluate the performance of different \gls{imu} sensors jointly with a multi-frequency, multi-constellation GNSS receiver for spoofing detection
    \item An evaluation of the proposed method in a real adversarial scenario, on our dedicated platform and a generic mobile phone to demonstrate the feasibility, practicality and limitations of our approach
\end{itemize}

After the related work in \cref{sec:related_work}, \cref{sec:sys-adv} discusses the system and adversary model including extended functionality available at the receiver. \cref{sec:methodology} presents the modifications of established methods we adopt to, (i) remove the limitations due to a needed known antenna mechanization model, and (ii) extend the statistical model to be agnostic of the relative position of the spoofer and the victim receiver. \cref{sec:experimental} discusses the experimental platform developed to test the modified statistical test. \cref{sec:evaluation} discusses the results and the achieved performance in spoofing detection for a static and mobile receiver, and the comparison between our dedicated platform and a commercial smartphone. \cref{sec:conclusion} concludes with possible future directions.

\section{Related Work}
\label{sec:related_work}
Detection of spoofing based on properties of the received signal is explored in \cite{WOS:000359380700146,WOS:000209006500003,WOS:000423143700015}. Changes in the acquisition matrix (e.g., the shape of the acquisition peak, number of peaks per acquisition channel) of the \gls{gnss} signal are generally good indicators of the presence of adversarial signals. While such an approach is highly effective, it requires direct access to the acquisition stage of the \gls{gnss} receiver, unavailable in commercial \gls{cots} receivers. Alternatives, such as \cite{DBLP:conf/ndss/SathayeLCR22}, use multiple channels to acquire separate peaks in the same acquisition space, with the drawback that reduced number of signals can be tracked at the same time. 

Transmission origin estimation based on the received signal power and on the receiver's \gls{agc} provide an indicative figure of the quality of the received signals \cite{WOS:000423143700015,akos2003,WOS:000209006500003}. However, changes in the \gls{agc} are often hard to relate to adversarial manipulation or variations in the environment of a mobile antenna (subject to time-varying multipath). Techniques generally referred to as \gls{sqm}, while providing immediate insight on the structure and quality of the \gls{gnss} signal quality, tend to perform poorly in a dynamic scenario. Similarly, metrics based on Doppler or pseudorange plausibility monitoring are effective and relatively low cost in their evaluation \cite{PapadimitratosJa:C:2008}, but can be thwarted by improved attacker hardware (e.g., more accurate clock distribution at the transmitter front-end) and better adversarial strategy (e.g., precise code phase alignment of the spoofed signals to the legitimate ones, \cite{SpangheroPP:C:2023}). 

In this context, two interesting recent improvements allowed more advanced spoofing countermeasures to be deployed in civilian COTS systems. First, inertial sensors improved in stability and accuracy even at the lower end of the segment as long as the integration time is short. Second, more feature-rich \gls{gnss} receivers are increasingly integrated in platforms providing additional sensors, computational power, and connectivity. 
This generally includes so-called raw measurements obtained by the \gls{gnss} receiver tracking loops and consists of the raw observables without any processing from the \gls{gnss} receiver's \gls{pnt} engine. Techniques based on validation of the Doppler shift of the received signal often allow detection of spoofed satellite signal, but the attacker can circumvent such detection using better and more stable reference sources at the adversarial transmitter \cite{PapadimitratosJ:C:2008}. Similarly, pseudorange measurement bounding also proved effective in detecting spoofed signals but with the limitation that often such detection system is dependent on a first acquisition in a benign scenario to establish a baseline \cite{PapadimitratosJa:C:2008,Jovanovic2014MultitestDA}. Such measurements are increasingly available even on mobile devices thanks to the Android Raw GNSS Measurements API, allowing hardening portable receivers \cite{WOS:000545000100015,10081330,Spensnavi.537}

Work on the GNSS-INS fusion shows that the current state of the inertial navigation quality is sufficient to improve the quality of GNSS-only measurements in a benign scenario, for a gamut of mobile platforms \cite{Sharma2021TimeSynchronizedGD,8915828,Lee2022GNSSFD}. Such devices can detect spoofing based on the inconsistency of the dynamics, e.g. when the spoofer causes rapid changes in the \gls{pnt} solution beyond the dynamics achievable by the mobile system \cite{Curran2017OnTU,Kujurnavi.629}. Hypothesis testing based on incongruities of the \gls{imu} measured acceleration and the \gls{pnt} provided by the \gls{gnss} receiver reliably detect spoofing attacks but do not provide any further information on (the complexity of) the attack, relying on the navigation processor outcome. A traditional approach relies on innovation testing while performing joint navigation and estimation using a Kalman filter (or other variations). While this greatly benefits robotics and autonomous systems, the improvements to navigation in \gls{gnss} denied conditions are limited, and low-cost \gls{imu}s cannot provide reliable inertial navigation. Ultimately, the strongest limitation is the quality of the \gls{imu} sensors used for recovering from \gls{gnss} spoofing and jamming, with \gls{imu} errors degrading the solution usually within a few minutes of the loss of \gls{gnss} lock. Accumulation of the integration error will grow in an unbound manner over time, making the innovation test result meaningless for anti-spoofing purposes. 
Also, integration window-based methods are generally slow in detecting an adversary as the innovation residual needs to increase beyond the confidence the filter has in the estimated covariance of the GNSS measurement. Practically, a subtle adversary slowly drifting the \gls{pnt} solution might not be detected until it causes major \gls{pnt} solution disruption.


Carrier phase measurements can provide considerable improvements to the quality and accuracy of the \gls{pnt} solution due to the much higher resolution of the carrier information, compared to code-based ranging. \gls{imu} measurements in tight GNSS-INS integration help resolve the integer ambiguity problem in differential \gls{gnss} systems where a joint baseline estimation with a reference station allows reliable spoofing detection even in a multipath-challenged environment (e.g., urban canyons). The main advantage consists in the dual robustness effect against the environment and potential attackers, but this requires external reference stations, limiting the applicability to scenarios where this is available. In the context ofthe  recent development of autonomous vehicular and aerial platforms, carrier phase measurements play a critical role in providing centimeter-level accurate positioning and enhancing spoofing countermeasures. As shown in \cite{WOS:000375213003001,Clements2022CarrierphaseAI,Hu_Bian_Ji_Li_2018,WOS:000356331204003}, the high resolution of the carrier phase information can be evaluated against high-frequency antenna motion to detect adversarial signals originating from a single transmitter. 
Specifically, high-frequency antenna motion can be leveraged to detect spoofed satellite signals \cite{WOS:000375213003001}, specifically in the case where the antenna dynamics are unidirectional and can be determined by a mechanization model. The latter can be complex to extract for moving antennas, where the amplitude and frequency of the motion can be arbitrary and multi-directional in space.

\section{System and adversary model}
\label{sec:sys-adv}
\textit{Adversary model - } Due to the open structure of civilian signals, the modulation, data content, frequency allocation, and signal parameters are known to the adversary for all civilian constellations. Hence, the adversary can use simulation, replay, relay, or adopt a combination of multiple methods, to generate signals that are valid from a physical layer and data content perspective and achieve the intended adversarial effect on the victim. Practically, we do not limit the attacker method to control the victim receiver, but if cryptographically enhanced signals are used, the attacker cannot modify any of the authenticated information and is limited to replay/relay of the secure blocks.

We assume the adversary transmits the spoofing signals from a single antenna. While an adversary could deploy multiple, synchronized transmitter nodes/antennas, the complexity of the attack would increase considerably. To achieve the correct spatial distribution of the spoofing signal in relation to the legitimate constellation, the adversary would need to place the transmitters in \gls{los} path to the victim and the legitimate satellite. Also, 
although possible, it is extremely challenging for the adversary to keep a tight synchronization among the transmitters over large distances and use enough transmitters to replicate the real carrier phase spreading. 

There is no limitation to the relative distance and position of the attacker and the victim as the attacker can position itself to maximize the chances of success. However, the attacker accuracy in tracking the victim receiver actual position is limited. It is generally unfeasible for the attacker to know with centimeter-level accuracy the position of the antenna phase center, which is a requirement to launch a stealthy spoofing overtake with carrier phase coherence. This limitation is valid in particular for mobile platforms, where the unpredictability of the victim movements makes the generation of carrier-phase locked spoofing signal unrealistic \cite{WOS:000375213003001,Jiadong2019}. Specifically, even if the attacker could potentially replicate a realistic carrier phase offset of the real constellation, it would not be able to track accurately enough a fast moving victim. 

\textit{System model -} A commercial, off-the-shelf GNSS receiver supporting multi-frequency and multi-constellation reception coupled with a commercial grade \gls{imu} sensor. Specifically, the \gls{gnss} receiver must provide raw measurements from the receiver tracking loops, in all constellations and frequencies the receiver is interested in monitoring. Access to the receiver own \gls{pnt} is also beneficial but the user can implement its own \gls{pnt} engine based on the raw measurements provided by the receiver.
The GNSS+IMU receiver provides raw, synchronous measurements from all sensors, without any further fusion to the processing system and with a known rigid transformation between the reference frames of the \gls{gnss} antenna phase center and the \gls{imu}. We do not restrict the mobility of the receiver, which can be static or mobile with different types of dynamics. On the other hand, we require that the antenna movement be characterized by two main components: low-frequency and high-frequency components. The first can be used for navigation in a canonical sensor fusion component, jointly with the \gls{gnss} \gls{pnt}. The second is usually filtered out for navigation purposes, but within the scope of this work, it is required to perform spoofing detection and mitigation. The level of dynamics must be high enough so that the platform can detect some movement of the \gls{gnss} antenna. A simplified view of the setup is given in  \cref{fig:simplified_setup}.

\section{Methodology}
\label{sec:methodology}

For a generic \gls{gnss} receiver, the satellite signal carrier phase depends on the satellite geometric distance and any atmospheric deviation. This is true more so that legitimate signals are transmitted from geometrically diverse points, corresponding to the true locations of the satellites. The observation at the receiver $r$ of the carrier phase for a generic satellite $s$ at range $\rho$ and time $t$ is defined in \cref{eqv:carrier-phase}, in accordance to \cite{Meurer2017}. The carrier-phase measurements are subject to clock offsets, $dt$, between the satellite and receiver compared to the constellation reference and phase delays in the instrumentation, $\phi_{r,j}$, $\phi^s_j$. $I^s_{r,j}$, $T^s_{r,j}$ are the ionospheric and tropospheric delays. 

\begin{equation}
    \label{eqv:carrier-phase}
    \begin{split} 
    \phi^s_{r}(t) = \frac{1}{\lambda} \rho^s_{r}(t) + (\phi_{r} - \phi^s) + \frac{c}{\lambda}(dt_r(t) - dt^s(t)) - I^s_{r}(t) + T^s_{r}(t) + N^s_{r} + n^s_{r,\phi}(t)
    \end{split}
\end{equation}

While the number of full phase cycles can be estimated using different techniques (e.g., \cite{navipedia_carrier_fixing}), the variation of carrier phase can be accurately measured by the receiver tracking loop by calculating the difference between the \gls{nco}-provided local copy of the carrier after aligning it to the satellite transmitted one. If the distance between the satellite and the receiver changes by more than one phase cycle ($\approx$~\SI{20}{\centi\meter} for GPS L1) the integer counter is updated to provide continuous tracking. 

\begin{figure}
    \centering
    \begin{subfigure}{.3\linewidth}
        \includegraphics[width=\linewidth]{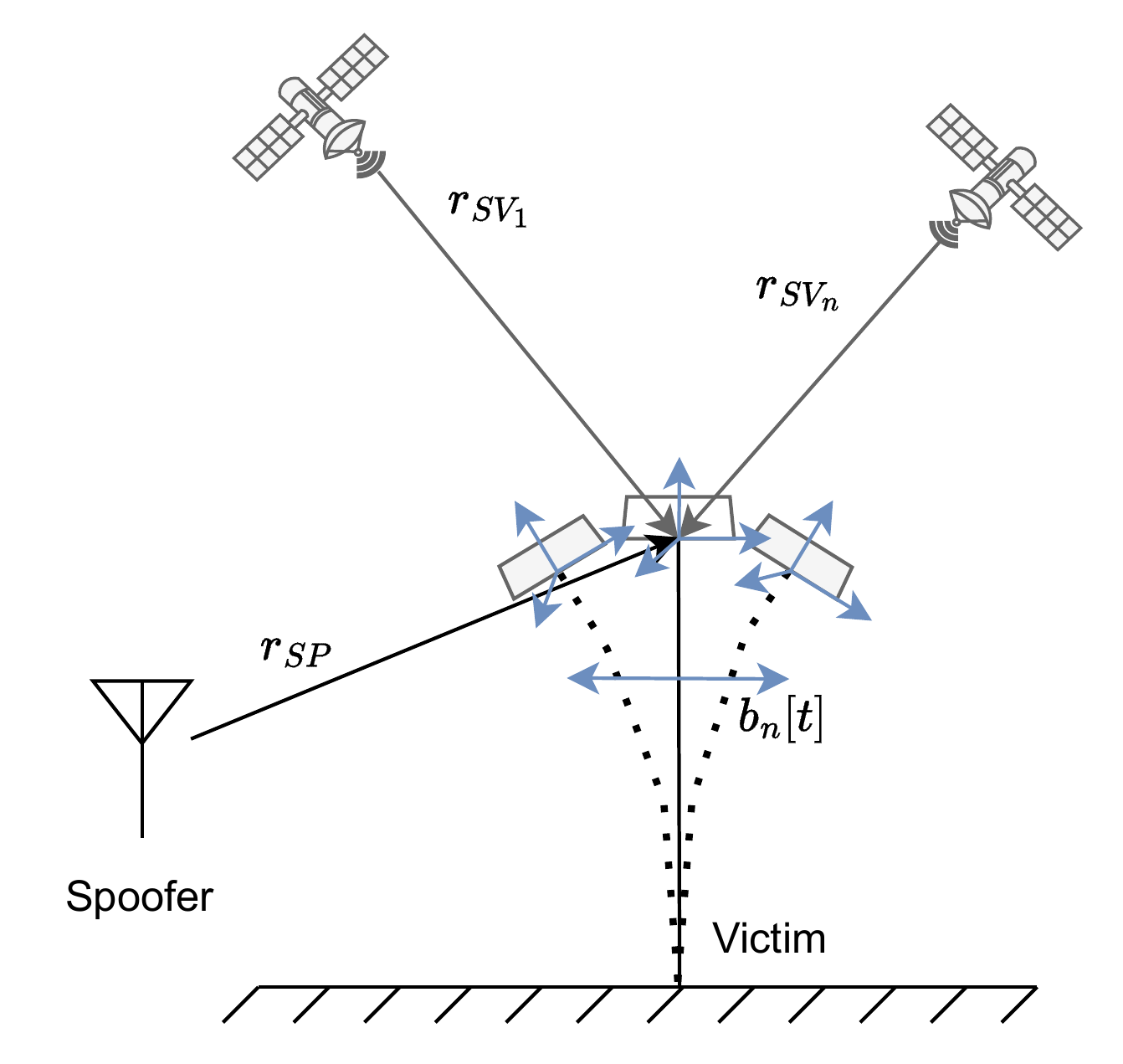}
        \caption{Simplified setup, with legitimate satellites in view with \gls{los} vector $r_{SV_n}$ and one adversarial transmitter with \gls{los} vector $r_{sp}$. }
        \label{fig:simplified_setup}
    \end{subfigure}
    \begin{subfigure}{.3\linewidth}
            \centering
            \includegraphics[width=\linewidth]{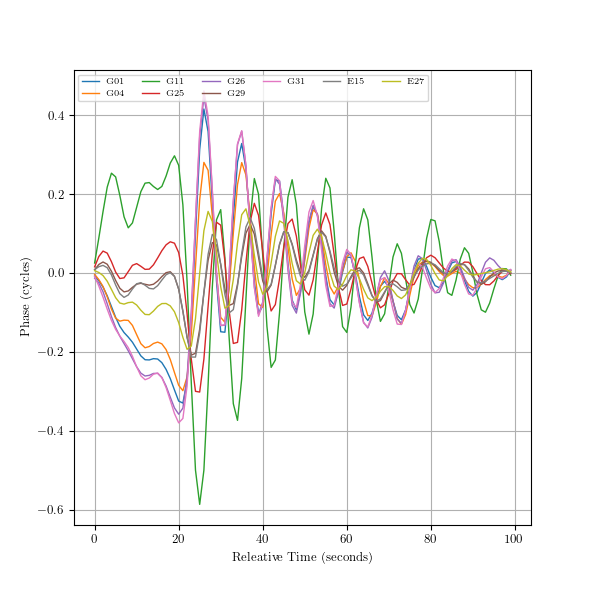}
            \caption{Carrier phase measurements in benign scenario. The spreading of the carrier phases is due to the satellite geometry.}
            \label{fig:carrier_meas_clean}
    \end{subfigure}
    \begin{subfigure}{.3\linewidth}
            \centering
            \includegraphics[width=\linewidth]{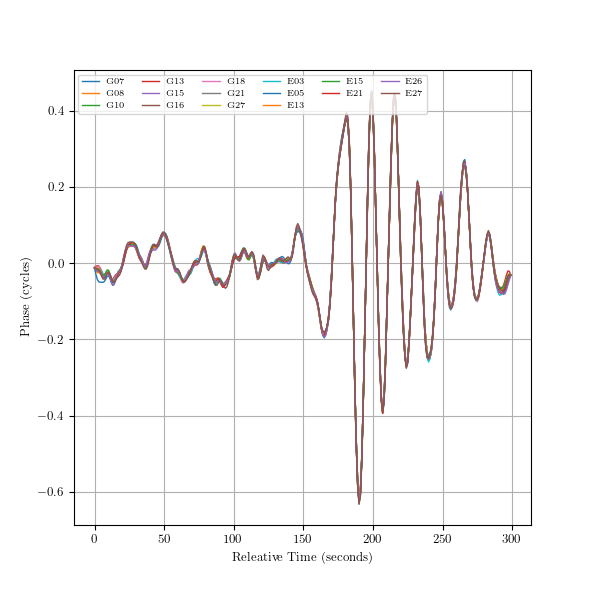}
            \caption{Carrier phase measurements in spoofed scenario. The same source of origin causes the carrier spreading to collapse.}
            \label{fig:carrier_meas_spoofed}
    \end{subfigure}
    \caption{System under consideration and example carrier measurements.}
\end{figure}

In a benign setting, the geometrical diversity directly influences each carrier phase measurement because of the different transmission positions. This effect is shown in \cref{fig:carrier_meas_clean}, justified by the carrier phase model in \cref{eqv:carrier-phase}. Similarly, \cref{fig:carrier_meas_spoofed} shows a subset of spoofed satellite signals transmitted by a single adversarial antenna. As all of these signals have the same propagation path, the carrier phase spreading collapses, with a reduced variance due to multipath effects.
It is worth noting that the carrier phase in \cref{fig:carrier_meas_clean,fig:carrier_meas_spoofed} is detrended (removing effects due to the satellite movement) for visualization purposes; practically this is not required by the method, which only operates on the high frequency components of the carrier phase. 

We are interested in modeling the carrier phase estimated at the receiver's \gls{pll} at time $t$, defined as $\Phi_{t_k}$, where the \gls{nco} smoothly tracks the signal carrier phase. 
Similarly to \cite{WOS:000375213003001}, two separate models are used for the carrier phase, in non-spoofing and spoofing conditions, considering the articulation of the receiver antenna.

The receiver antenna position is defined as $\hat{b}[k]$. This is obtained by applying a high pass filter to the \gls{imu} linear acceleration and pose, and subsequent integration obtained of the $\hat{v}[k],\hat{p}[k]$ velocity and position estimates relative to the antenna reference frame. The exact cut-off frequency of the high pass filter is not fundamental in this process, as it is only used to remove the sensor bias that would cause a diverging integral solution of the displacement. While bias and drift in the \gls{imu} are the main contributor of noise over long integration periods, these can be removed over very short integration periods especially as we are interested in the high frequency components of the antenna movement.  The estimation of the movement displacement is done by double integration of the linear acceleration obtained from the \gls{imu} after removing the effect of gravity. The process follows a state of the art probabilistic approach as described in \cite{Madgwick2010AnEO}.

Based on \cref{eqv:carrier-phase}, the geometrical distance, $\rho$, between the \gls{gnss} receiver and the any satellite in view with valid phase measurement at a discrete-time $k$ is first expressed as a function of the satellite-receiver \gls{los} vector. Then, the resulting estimate is rotated from the \gls{ecef} reference frame, efficiently estimating the satellite position in a local relative frame where the motion vector of the antenna is defined. As the antenna displacement is small in comparison to the physical distance between the satellite and the receiver, the final form of the carrier phase estimate is given in \cref{eqv:carrier-phase-first-order}. 

\begin{equation}
    \label{eqv:carrier-phase-first-order}
    \begin{split} 
    \phi^j[k] \approx \frac{1}{\lambda} (\sqrt{(\bm{r}^j[k])^T\bm{r}^j[k]} + (\bm{\hat{r}}^j)^TA^T\bm{b}_n[k]) + (\phi^j_{r} - \phi^j_s) + \frac{c}{\lambda}(dt_r[k] - dt^s[k]) - I^j[k] + T^j[k] + N^j + n^j_{\phi}[k]
    \end{split}
\end{equation}

The high-frequency components are extracted from \cref{eqv:carrier-phase-first-order}, and are used in the hypothesis test. The low-frequency carrier phase component can be approximated with a polynomial interpolation with fixed coefficients $\beta_{1..3}$, as shown in \cref{eqv:carrier-phase-LF},  which will then be minimized by fitting the carrier phase model to the measurements. The measured carrier phase must be continuous and connected during the window under test. Cycle slips in the carrier phase make the specific data window unusable by our method or need to be addressed before so that the carrier phase model in \cref{eqv:carrier-phase} applies. Additionally, compared to \cite{WOS:000375213003001}, the corrections of the carrier slip needs to be aware of the system dynamics and take into account the \gls{imu} measured antenna displacement so that the consistency between the actual antenna displacement, and the repaired carrier phase is maintained. 

\begin{equation}
    \label{eqv:carrier-phase-LF}
    \begin{split} 
    \phi^j_{LF}[k] = \frac{1}{\lambda} \sqrt{(\bm{r}^j[k])^T\bm{r}^j[k]} +  (\phi^j_{r} - \phi^j_s) + \frac{c}{\lambda}(dt_r[k] - dt^s[k]) - I^j[k] + T^j[k] + N^j  \approx \beta^j_0 + \beta^j_1[k - k_0] + \frac{1}{2}\beta^j_2[k - k_0]^2
    \end{split}
\end{equation}

\textit{Carrier phase models -} By combining \cref{eqv:carrier-phase-first-order,eqv:carrier-phase-LF}, the carrier phase estimate for a legitimate scenario, referenced to the antenna local frame for each satellite to receiver \gls{los} vector is obtained, as shown in \cref{eqv:non-spoofed-carrier-phase}. The same method is used for the spoofed case shown in \cref{eqv:spoofed-carrier-phase}, with the antenna to satellite \gls{los} vector replaced by an unknown vector $\hat{r}_{SP}$, the \gls{los} vector between the victim antenna and the spoofer transmitting antenna, as shown in \cref{fig:simplified_setup}. The expression of the legitimate and spoofed carrier phase are identical, but for the \gls{los} vector which in one case points to the legitimate satellite and in the other to the spoofing transmitter. 

\begin{multicols}{2}
\noindent
\begin{equation}
    \label{eqv:non-spoofed-carrier-phase}
        \begin{split} 
        \phi^j[k] \approx \frac{1}{\lambda} (\bm{\hat{r}}^j)^TA^T\bm{b}_n[k] + \beta^j_0 + \beta^j_1[k - k_0] + \\ \frac{1}{2}\beta^j_2[k - k_0]^2 + n^j_{\phi}[k]
        \end{split}
\end{equation}
\columnbreak
\begin{equation}
    \label{eqv:spoofed-carrier-phase}
        \begin{split} 
         \phi^{sp}[k] \approx \frac{1}{\lambda} (\bm{\hat{r}}^{sp})^TA^T\bm{b}_n[k] + \beta^j_0 + \beta^j_1[k - k_0] + \\ \frac{1}{2}\beta^j_2[k - k_0]^2 + n^j_{\phi}[k]
         \end{split}
\end{equation}
\end{multicols}

\textcolor{black}{One limitation due to commercially available receivers, whose structure is unknown, is that the raw carrier measurements from the tracking loops are essentially provided by a black box. \cite{WOS:000375213003001} states that sampling the carrier phase at the center of the receiver coherent integration window increases the \gls{snr} of the beat carrier estimate, as it whitens the noise figure of the estimate. Unfortunately, without modifications or knowledge of the receiver structure, it is impossible to know at which point the observables are measured, so we cannot operate under the assumption that the estimate noise is white. Nevertheless, we will empirically show that this assumption can be removed while minimally affecting our method. Second, as the \gls{nco} tracks the carrier frequency using a \gls{pll}, the \gls{pll} bandwidth determines the maximum dynamics of the receiver (e.g., practically limiting the acceleration of the antenna before carrier phase tracking is lost).}

Spoofing detection is performed by a statistical test similar to \cite{WOS:000375213003001}, where the null hypothesis is the benign scenario and the alternative is the spoofed case. The ratio of the likelihood between the two distributions is our decision metric, as shown in \cref{eqv:likelihood_decision}. The threshold, $c$, can be determined dynamically based on the quality of the fit for each distribution, but for simplicity, we use a static threshold. While the statistical test is the same, the actual distributions depend on the specific antenna dynamics. 

Here and in the following sections we use a simplified notation for the spoofed and legitimate carrier phase expression, using a unified receiver-transmitter vector notation $\bm{\hat{r}_x}$ (specificity will be added when necessary).

\begin{equation}
\label{eqv:likelihood_decision}
    \Gamma = \frac{\mathcal{L}(H_0|x)}{\mathcal{L}(H_1 | x)}
    \\
    \begin{cases}
    \text{if\ } \Gamma > c \text{,\ do not reject\ } H_0\\
    \text{if\ } \Gamma < c \text{,\ reject\ } H_0\\
    \text{if\ } \Gamma = c \text{,\ reject\ } H_0 \text{\ with probability\ } q\\
    \end{cases}
\end{equation}

\textit{Estimation of the antenna displacement -} Compared to previous approaches, where the attitude of the victim antenna is unknown or determined by other external mechanical measurements, our setup relies on the \gls{imu} to first estimate the movement and displacement of the victim antenna, before applying the decision statistics. 
The direction of motion, $\bm{\hat{r_a}} = A^T\bm{\hat{b}}$, is obtained by the \gls{imu} integration. The direction of motion is calculated as unit vectors, projected in the local reference frame.
The motion amplitude, $p[k]$, of the $\bm{b}^T_n[k]$ vector is defined as the norm of the motion vector based on the \gls{imu} integration, with the sign based on the angle between the motion vector and the estimated unit direction. The integration timescale of $p[k]$ is given by the sampling rate of the \gls{imu}, which is generally few orders of magnitude higher than the \gls{gnss} to allow accurate tracking during the high-frequency motion.
\begin{equation}
    \label{eqv:rho}
    p[k] = -sign \left( \frac{{\bm{b}_n} \cdot \bm{\hat{r}_a}}{||{\bm{b}_n|| ||\bm{\hat{r}_a}}||} \right) ||\bm{{b}}_n||
\end{equation}

Given that we do not know the real position of the victim antenna (during spoofing at least, but it is not a requirement in general), the translation between the local frame of the antenna and the global frame (e.g., ECEF) is unknown. We solve this by relying on the \gls{imu} measurements to identify the directions of movement and by obtaining the attitude in the global frame by applying a \gls{mle}. The \gls{mle} estimator does not calculate the real transformation but instead allows retrieval of an unity attitude vector that is parallel to the one in the global frame.

If the carrier phase model in \cref{eqv:non-spoofed-carrier-phase,eqv:spoofed-carrier-phase} is used without further simplifications, the high-pass filtering can instead be defined as \cref{eqv:system-of-equation}, and the QR-factorization is performed on the normalized version of the now N-by-6 matrix on the right-hand side of \cref{eqv:system-of-equation}.

\begin{equation}
\label{eqv:system-of-equation}
    \begin{bmatrix} \phi^j_k\\ \phi^j_{k-1} \\ \phi^j_{k-2} \\ \phi^j_{k-3} \\ \vdots  \\ \phi^j_{k-N}
    \end{bmatrix}
    =
    \begin{bmatrix}
        1 & 0 & 0 & \frac{1}{\lambda} \bm{b}^T_{n_{k}} \\
        1 & 1 & \frac{1}{2} \cdot 1^2 & \frac{1}{\lambda} \bm{b}^T_{n_{k-1}} \\
        1 & 2 & \frac{1}{2} \cdot 2^2 & \frac{1}{\lambda} \bm{b}^T_{n{k-2}} \\
        1 & 3 & \frac{1}{2} \cdot 3^2 & \frac{1}{\lambda} \bm{b}^T_{n{k-3}} \\
        \vdots & \vdots & \vdots & \vdots \\
        1 & N & \frac{1}{2} \cdot N^2 & \frac{1}{\lambda} \bm{b}^T_{n{k-N}}
    \end{bmatrix}
    \begin{bmatrix}
    \beta^j_0 \\ \beta^j_1 \\ \beta^j_2 \\ A\bm{\hat{r}}^x
    \end{bmatrix}
    +
    \begin{bmatrix}
         n^j_{\phi_{k}} \\ n^j_{\phi_{k-1}} \\ n^j_{\phi_{k-2}} \\ n^j_{\phi_{k-3}} \\ \vdots  \\ n^j_{\phi_{k-N}}
    \end{bmatrix}
\end{equation}

At this stage, the $\beta_{1..3}$ coefficients for the \gls{los} vector to each satellite need to be minimized with an optimization method. A QR-factorization on the system of $N$ equations, where each line corresponds to a measurement during the sampling window $[k, k-N]$, returns a high-pass filtered version of the carrier phase (as a result of the factorization). This is repeated for each satellite $j$ that is in view and for which we have valid carrier measurements. The Q matrix in the QR-factorization returns the quantity of interest, which is the filtered carrier phase and noise vectors shown in \cref{eqv:norm_phi,eqv:norm_noise}.

\begin{multicols}{2}
\noindent
\begin{equation}
   \label{eqv:norm_phi}
    \begin{split}
    \begin{bmatrix} z^j[k] \\ z^j[k-1] \\ z^j[k-2] \\ z^j[k-3] \\ \vdots  \\ z^j[k-N]
    \end{bmatrix} = \frac{1}{\sigma^j}(Q^j)^T
        \begin{bmatrix} \phi^j[k] \\ \phi^j[k-1] \\ \phi^j[k-2] \\ \phi^j[k-3] \\ \vdots  \\ \phi^j[k-N]
    \end{bmatrix}
    \end{split}
\end{equation}\columnbreak
\begin{equation}
\label{eqv:norm_noise}
\begin{split}
    \begin{bmatrix} \eta^j[k] \\ \eta^j[k-1] \\ \eta^j[k-2] \\ \eta^j[k-3] \\ \vdots  \\ \eta^j[k-N]
    \end{bmatrix} = \frac{1}{\sigma^j}(Q^j)^T
        \begin{bmatrix} n_\phi^j[k] \\ n_\phi^j[k-1] \\ n_\phi^j[k-2] \\ n_\phi^j[k-3] \\ \vdots  \\ n_\phi^j[k-N]
    \end{bmatrix}
    \end{split}
\end{equation}
\end{multicols}

With the same derivations as in \cite{WOS:000375213003001}, by applying a Least-Square estimation on \cref{eqv:system-of-equation} with the normalized carrier phase and noise vectors from \cref{eqv:norm_phi,eqv:norm_noise} we obtain \cref{eqv:QRSolution-3D}, which is the expression of the dynamics carrier phase model where $\bm{R}$ is obtained by the QR-factorization. Here, the first Eq. (1,3) equations in the system only pertain to the antenna's own motion and can be integrated independently of the test hypothesis, leading to the same quantities. Similarly, Eq. (6,$N$) are identical in either hypothesis case as they are all simplified by the QR-factorization.

\begin{equation}
\label{eqv:QRSolution-3D}
    \begin{bmatrix} 
     z^j[k] \\ z^j[k-1] \\ z^j[k-2]  \\ \vdots  \\ z^j[k-N]
    \end{bmatrix}
    =
    \begin{bmatrix}
        \bm{R}_{6x6} \\
        \bm{0}_{6xN}
    \end{bmatrix}
    \begin{bmatrix}
    \beta^j_0 \\ \beta^j_1 \\ \beta^j_2 \\ 
    A\bm{\hat{r}}^x
    \end{bmatrix}
     +
    \begin{bmatrix} \eta_\phi^j[k] \\ \eta_\phi^j[k-1] \\ \eta_\phi^j[k-2] \\ \eta_\phi^j[k-3] \\ \vdots  \\ \eta_\phi^j[k-N]
    \end{bmatrix}
\end{equation}

\textit{Probability distributions} - The remaining Eq. (3,5) are the ones of interest (\cref{eqv:QRoutput-3D}) and represent the basis for our Neyman-Pearson test whose statistics are defined in \cref{eqv:likelihoodNonSp} and \cref{eqv:likelihoodSp} for the non-spoofed and spoofed hypothesis respectively.

\begin{equation}
\label{eqv:QRoutput-3D}
    \begin{bmatrix} z^j[k-3] \\ z^j[k-4] \\ z^j[k-5]
    \end{bmatrix} =
    \begin{bmatrix} 
    R^j_{44} & R^j_{45} & R^j_{46} \\
    0 & R^j_{55} & R^j_{56} \\
    0 & 0 & R^j_{66} \\
    \end{bmatrix}
    A\bm{\hat{r}}^x + 
    \begin{bmatrix} \eta^j[k-3] \\ \eta^j[k-4] \\ \eta^j[k-5]
    \end{bmatrix}
\end{equation}

\begin{multicols}{2}
\noindent
    \begin{equation}
        \begin{split}
         \label{eqv:likelihoodNonSp}
            \mathcal{L}(A, H_0 | \bm{z}^1,\dots, \bm{z}^L)= w\text{exp}(-\frac{1}{2} \sum_{j=1}^L&[R^jA\bm{\hat{r}}^j - \bm{z}^j]^T \\ &\cdot [R^jA\bm{\hat{r}}^j - \bm{z}^j])
        \end{split}
    \end{equation}
    \columnbreak
\begin{equation}
        \begin{split}
        \label{eqv:likelihoodSp}
             \mathcal{L}(\bm{\hat{r}}^{sp}, H_1 | \bm{z}^1,\dots, \bm{z}^L) = w\text{exp}(-\frac{1}{2} \sum_{j=1}^L&[R^j\bm{\hat{r}}^{sp} - \bm{z}^j]^T \\ &\cdot [R^j\bm{\hat{r}}^{sp} - \bm{z}^j])
        \end{split}
    \end{equation}
\end{multicols}

To have a complete formulation of the hypothesis test, the system optimizes, based on the displacement vector estimation $\bm{b}^T_n[k]$, the indicator vectors for the \gls{los} vector between the receiver and the legitimate satellites as shown in \cref{eqv:Jnonsp_opt}. The actual transformation is still unknown, but the optimization process maximizes the likelihood of the \gls{los} direction, shown in \cref{eqv:Jnonsp_opt}. In the first case, we estimate $R^jA\bm{\hat{r}^j}$ for each antenna-satellite \gls{los} vector. In the second case, we perform the same estimation for the victim-spoofer \gls{los} vector. 
Here, the advantage of a strap-down \gls{imu} to track the antenna position is clear. The dynamics of the antenna in the benign and spoofed case can be directly extracted from the local \gls{imu}, simplifying the determination of the unknown antenna movement.

\begin{equation}
\label{eqv:Jnonsp_opt}
\begin{split}
     &\text{Find\ } A ~ \text{subject to\ } A^TA = I\\
     &\begin{split}
        \text{to minimize:\ } J_{nonsp}(A) = \frac{1}{2} \sum_{j=1}^L&[R^jA\bm{\hat{r}}^j - \bm{z}^j]^T \\ &\cdot [R^jA\bm{\hat{r}}^j - \bm{z}^j] 
     \end{split}\\     
\end{split}
\end{equation}

A similar optimization problem is solved to find the $\bm{\hat{r}}^{sp}$ that maximizes the likelihood for the spoofer-victim \gls{los} vector, with the same optimization problem as stated in \cref{eqv:Jnonsp_opt}.
The optimized values for $\bm{\hat{r}}^{sp}_{opt}$ and $\bm{A}_{opt}$ are used in the final formulations of the probability distributions in \cref{eqv:likelihoodNonSp,eqv:likelihoodSp} to obtain the decision statistics \cref{eqv:decision_test}.


\textit{Decision statistics and metrics} - The decision statistics are implemented as in \cref{eqv:likelihood_decision}, but by evaluating the negative log-likelihood of the \cref{eqv:likelihoodNonSp,eqv:likelihoodSp} within \cref{eqv:likelihood_decision}, which in result gives \cref{eqv:decision_test}, as in \cite{WOS:000375213003001}.

\begin{equation}
    \gamma = J_{sp}(\bm{\hat{r}_{opt}}) - J_{nonsp}(\bm{A}_{opt})
    \label{eqv:decision_test}
\end{equation}

Differently from \cite{WOS:000375213003001}, the detector threshold is fixed and set to 0, obtained by simplifying the expression of $c$ in \cref{eqv:likelihood_decision} While this can be optimized to minimize the false positive rate, it proved to contribute minimal improvement compared to a simple positive/negative decision. 
The test is performed for each interval where there is sufficient movement to generate the required high frequency oscillations we base the detection on. We define each of the sampling windows where this is possible as an \emph{event}. The test is conclusive and with a determined outcome only if the overall instantaneous acceleration of the device (calculated as the $L_1$-norm of the acceleration values) is above an experimentally determined threshold, as defined in \cref{sec:evaluation}. In all other cases, the event is considered inconclusive, as there is not enough movement in the antenna to execute the spoofing detection mechanism.


\section{Experimental setup}
\label{sec:experimental}





\begin{figure}
    \centering
        \begin{subfigure}[t]{0.3\linewidth}
            \centering
            \includegraphics[width=\linewidth]{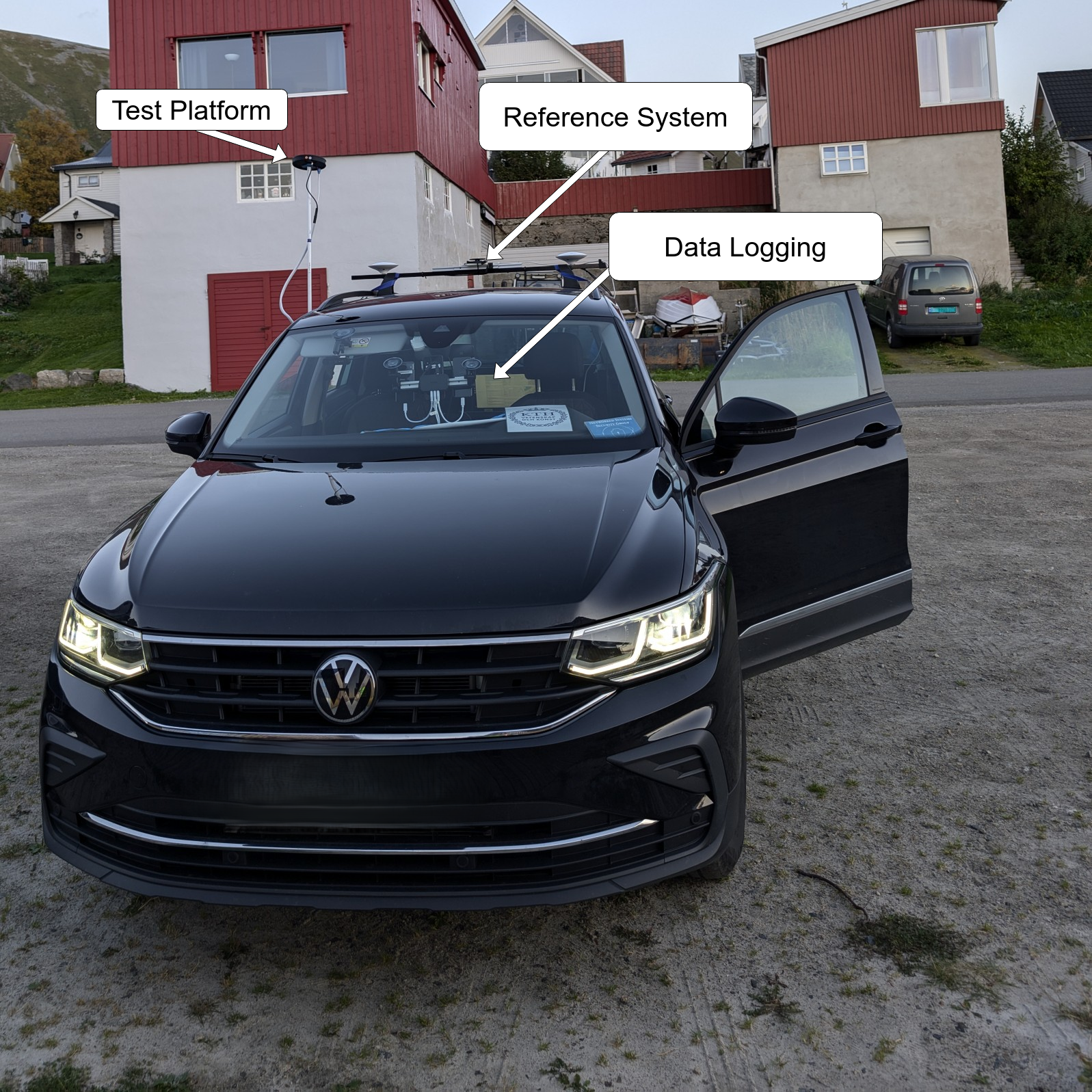}
            \caption{Dynamic tests fixture: the antenna system is mounted on the vehicle. Two configurations are used, either on a flexible pole or strap-down to the vehicle}
            \label{fig:dynamic-test-rig}
        \end{subfigure}
        \begin{subfigure}[t]{0.3\linewidth}
            \centering
            \includegraphics[width=\linewidth]{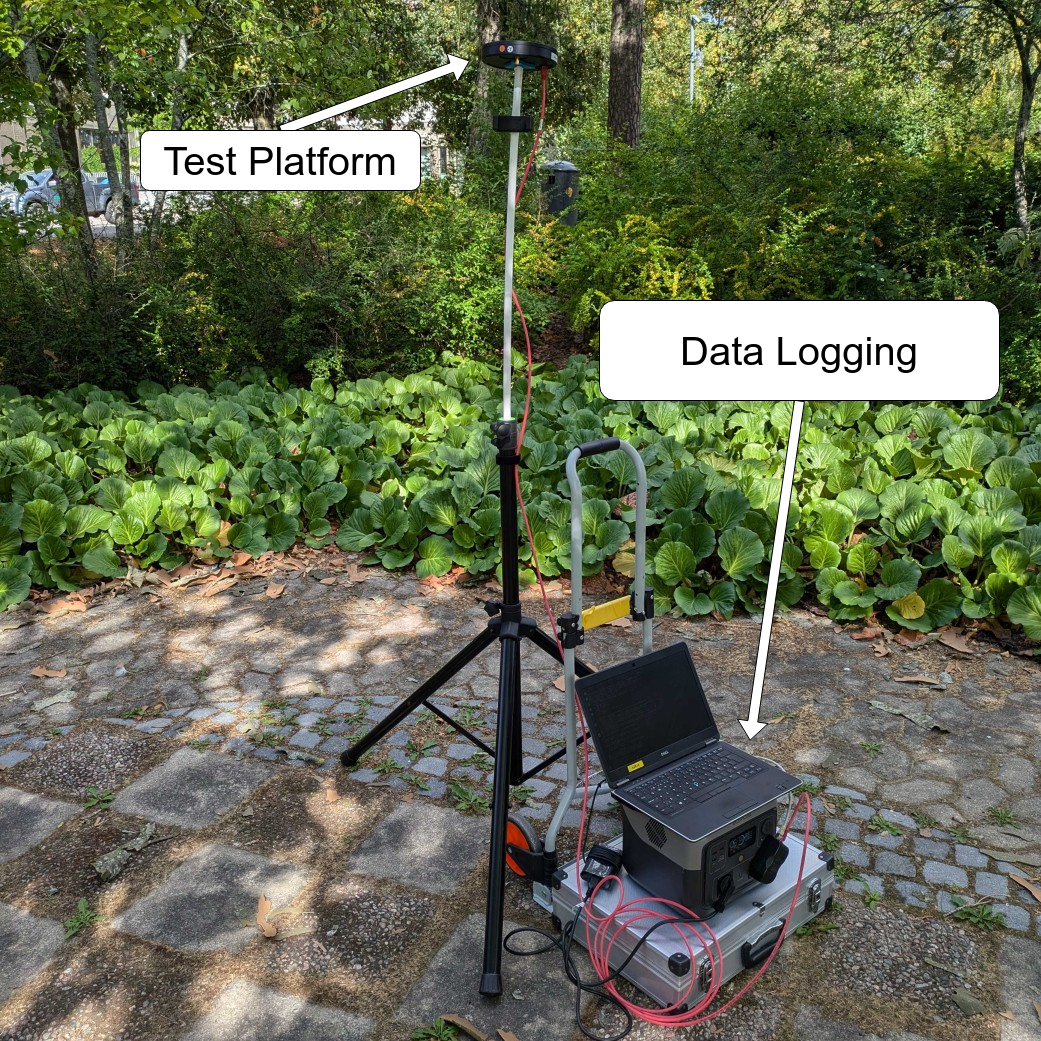}
            \caption{Static tests fixture: the antenna can move freely in all directions.}
            \label{fig:static-test-rig}
        \end{subfigure}
        \begin{subfigure}[t]{0.3\linewidth}
            \centering
            \includegraphics[width=\linewidth]{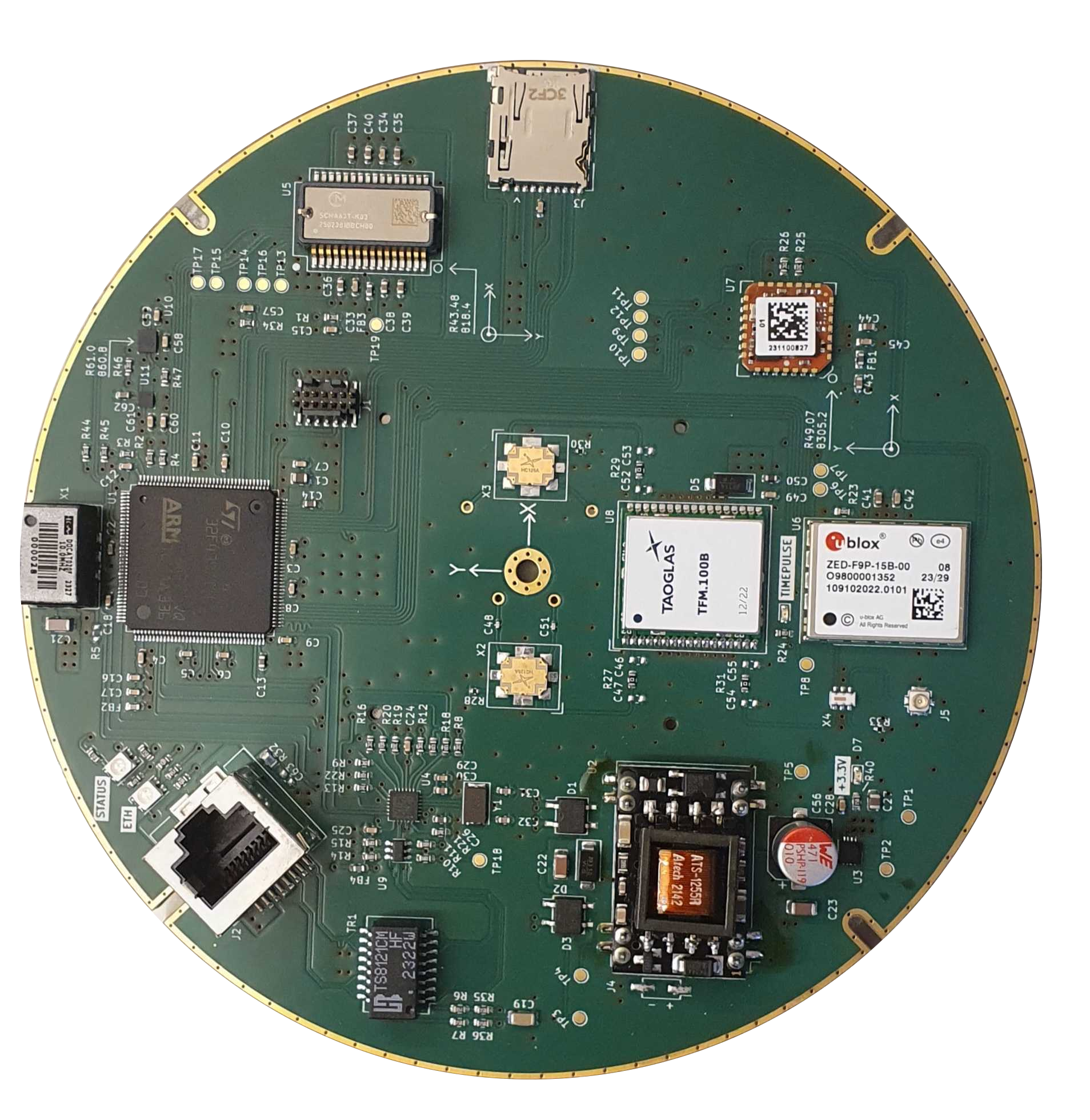}
            \caption{Data acquisition platform with multiple INS sensors and a precision GNSS receiver}
            \label{fig:platform}
        \end{subfigure}
        \caption{Test setups and acquisition platforms}
\end{figure}

\textit{Measurement device - } The platform designed to evaluate the method presented here has an U-Blox F9P dual band L1/L5 quad-constellation \gls{gnss} receiver \cite{ubloxf9p}. The platform is designed to allow testing of the same components commonly found on a modern GNSS-equipped mobile phone in a controlled and repeatable environment. The GPDF6010.A all-band high precision \gls{gnss} stacked patch antenna integrated with the platform uses a matched TAOGLASS-TFM-100B amplifier frontend as signal conditioner. Additionally, an external \gls{gnss} receiver or recorder can be connected to the antenna port for direct comparison with other \gls{gnss} measurement systems or raw baseband sampling. The sampling rate of the sensors and the \gls{gnss} receiver can be adjusted based on the application requirements. The device provides on-board computation and storage, mostly used for initialization tasks and sampling. An overview of the acquisition device is shown in \cref{fig:platform}.

\begin{figure}
  \centering
    \begin{subfigure}{0.3\linewidth}
    \centering
      \includegraphics[width=\linewidth]{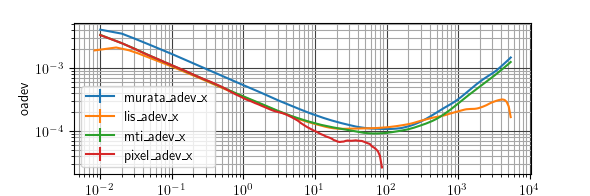}
      \caption{X axis}
    \end{subfigure}
    \begin{subfigure}{0.3\linewidth}
        \centering
      \includegraphics[width=\linewidth]{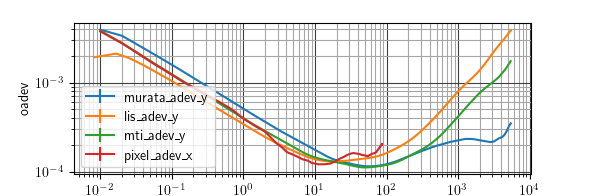}
      \caption{Y Axis}
    \end{subfigure}
    \begin{subfigure}{0.3\linewidth}
        \centering
      \includegraphics[width=\linewidth]{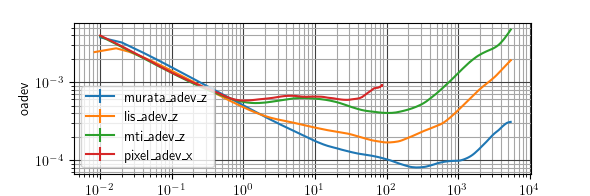}
      \caption{Z Axis}
    \end{subfigure}    
  \caption{Allan deviation for three different grade accelerometers}
  \label{fig:accelerometers_comp}
\end{figure}

\begin{figure}
  \centering
    \begin{subfigure}{0.3\linewidth}
    \centering
      \includegraphics[width=\linewidth]{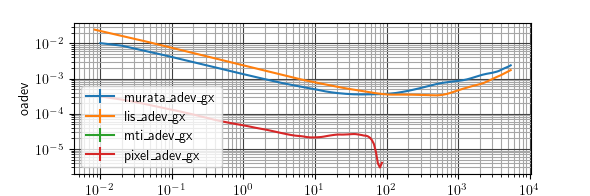}
      \caption{X axis}
    \end{subfigure}
    \begin{subfigure}{0.3\linewidth}
        \centering
      \includegraphics[width=\linewidth]{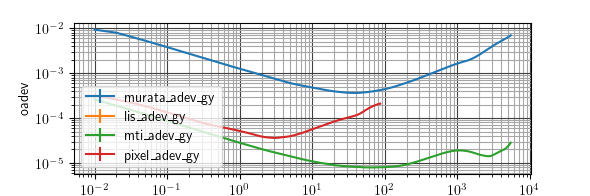}
      \caption{Y Axis}
    \end{subfigure}
    \begin{subfigure}{0.3\linewidth}
        \centering
      \includegraphics[width=\linewidth]{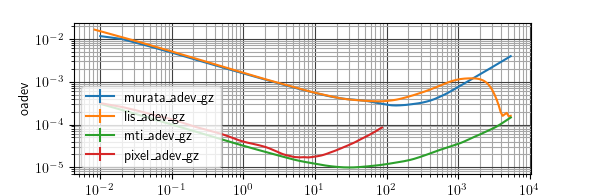}
      \caption{Z Axis}
    \end{subfigure}    
  \caption{Allan deviation for three different grade gyroscopes}
  \label{fig:gyro_comp}
\end{figure}

Three independent inertial sensors of different specifications, ranging from general purpose, mass production devices to advanced commercial \gls{imu} \gls{som} are embedded in the platform. The low-cost mass production \gls{imu} combines an ST Microelectronics LSM6DSV inertial sensor and ST Microelectronics LIS2MDL magnetometer, the mid-tier device is a Murata SCHA63T and the reference device is the Xsense \textit{MTI-3}, which also supports integrated sensor fusion. 
Calibration of each inertial platform is performed using an in-house test stand and a Ferraris calibration method \cite{Ferraris1994CalibrationStandards}. Notably, the SCHA63T is not coupled with a magnetometer, but we rely on the other available sensors for magnetic sensing. An overview of the declared performance of each sensor is provided in \cref{tab:sensor-specs}. \cref{fig:accelerometers_comp,fig:gyro_comp} shows the overlapping Allan Deviation \cite{Allan1987} per sensor axis as a quality measurement of the sensors. We remark that the SCHA63T sensor provides unprocessed sensor readings, in contrast to the MTI-3 and LSM6D devices, which both implement internal filtering and conditioning of the measurements. 

We also use a Google Pixel 8 with Android 14 to sample raw \gls{gnss} data in multi-constellation mode along with sensor data comprising of 3D acceleration, angular rate, and magnetometer. Further investigation shows that the Pixel 8 mobile phone uses a TDK ICM45631 accelerometer and gyroscope, combined with a Memsic MMC56X3X magnetometer. The \gls{gnss} receiver model used in the Pixel 8 phone is part of the Tensor G3 chipset. Further information regarding the specific capabilities of the chipset is unfortunately unknown.

Static tests are performed using the test stand in \cref{fig:static-test-rig}, with the antenna mounted on a flexible beam that allows movement in all directions with a predictable dampening action. Tests with mobility are performed using a vehicle where the antenna is mounted on a flexible mast similarly to \cref{fig:static-test-rig}, so that the oscillation is combined with the actual car movement, as shown in \cref{fig:dynamic-test-rig}. 

\textit{Tests - } Testing is performed in various scenarios, both static and mobile. Validation of the results is performed in three scenarios: benign, adversarial and mixed. 
Adversarial-only static tests are performed without causing any disturbance as the transmission is performed within a protected environment. Due to the strict limitations for transmission in the L-band, it is not possible to transmit over-the-air without the approval of the competent authority. The tests conducted in complete shielding from the real \gls{gnss} signals were designed to show the accuracy of our scheme in detecting spoofing. In all controlled reference cases, transmission of the spoofed signals is done using a single antenna, positioned in proximity to the victim. The distance between the victim receiver and the adversarial transmitter ranges between \SI{2}{\meter} and \SI{7}{\meter}, but it is not a limitation to either the attack or the countermeasure presented in this work. Similarly, benign-only static tests are performed in open sky, as a baseline for the non-adversarial case. For the benign scenario the chosen location is an urban setting, where several multistory buildings are present but no severe multipath is observed. A summary of the baseline test cases is provided in \cref{tab:experimental_benign_baseline,tab:experimental_spoofing_baseline} for the benign and spoofed cases respectively.

Realistic validation with over-the-air real and adversarial signals is performed at Jammertest 2024 \cite{jammertest-event}, where transmission of live spoofing signals is performed under the authority of NKOM, TestNor, and FFI among the other organizers. A comprehensive test suite is conducted over several days, including synchronous and asynchronous spoofing and meaconing. Tests are conducted in open sky, with the possibility of transitioning from benign to adversarial areas. In this setup, both static and mobile tests are conducted. A summary of the tests, data points, and settings used is given in \cref{tab:jammertest_testing}.

\begin{table}[!htbp]
\centering
\renewcommand{\arraystretch}{1.3}
\setlength{\tabcolsep}{7pt}
\resizebox{\textwidth}{!}{%
\begin{tabular}{|c|c|c|c|c|c|c|c|c|c|}
\hline
\textbf{Type} & \textbf{Subsystem} & \textbf{Range (\si{\gforce})} & \textbf{Bias (\si{\milli\gforce})} & \textbf{Noise (\si{\micro g\per\sqrt{\hertz}})} & \textbf{Range (\si{\degree\per\second})} & \textbf{Drift (\si{\degree\per\hour})} & \textbf{Noise (\si{\milli\degree\per\second\per\sqrt{\hertz}})} & \textbf{Range (\si{G})} & \textbf{Noise (\si{\milli G})} \\
\hline \hline
MTI-3-5A & Accelerometer & $\pm{16}$ & 0.03 & 120 & \cellcolor{gray!15} & \cellcolor{gray!15} & \cellcolor{gray!15} & \cellcolor{gray!15} & \cellcolor{gray!15} \\
\hline
SCHA63T & Accelerometer & $\pm{6}$ & 13.5 & 59.6 & \cellcolor{gray!15} & \cellcolor{gray!15} & \cellcolor{gray!15} & \cellcolor{gray!15} & \cellcolor{gray!15} \\
\hline
LSM6DSV & Accelerometer & $\pm{16}$ & 12 & 60 & \cellcolor{gray!15} & \cellcolor{gray!15} & \cellcolor{gray!15} & \cellcolor{gray!15} & \cellcolor{gray!15} \\
\hline
LSM6DSR & Accelerometer & $\pm{16}$ & 10 & 60 & \cellcolor{gray!15} & \cellcolor{gray!15} & \cellcolor{gray!15} & \cellcolor{gray!15} & \cellcolor{gray!15} \\
\hline \hline
MTI-3-5A & Gyroscope & \cellcolor{gray!15} & \cellcolor{gray!15} & \cellcolor{gray!15} & 2000 & 10 & 7 & \cellcolor{gray!15} & \cellcolor{gray!15} \\
\hline
SCHA63T & Gyroscope & \cellcolor{gray!15} & \cellcolor{gray!15} & \cellcolor{gray!15} & 300 & 1.64 & 15 & \cellcolor{gray!15} & \cellcolor{gray!15} \\
\hline
LSM6DSV & Gyroscope & \cellcolor{gray!15} & \cellcolor{gray!15} & \cellcolor{gray!15} & 4000 & 3600 & 2.8 & \cellcolor{gray!15} & \cellcolor{gray!15} \\
\hline
LSM6DSR & Gyroscope & \cellcolor{gray!15} & \cellcolor{gray!15} & \cellcolor{gray!15} & 4000 & 3600 & 5 & \cellcolor{gray!15} & \cellcolor{gray!15} \\
\hline \hline
MTI-3-5A & Magnetometer & \cellcolor{gray!15} & \cellcolor{gray!15} & \cellcolor{gray!15} & \cellcolor{gray!15} & \cellcolor{gray!15} & \cellcolor{gray!15} & 8 & 0.5 \\
\hline
SCHA63T & Magnetometer & \cellcolor{gray!15} & \cellcolor{gray!15} & \cellcolor{gray!15} & \cellcolor{gray!15} & \cellcolor{gray!15} & \cellcolor{gray!15} & N/A & N/A \\
\hline
LIS2MDL & Magnetometer & \cellcolor{gray!15} & \cellcolor{gray!15} & \cellcolor{gray!15} & \cellcolor{gray!15} & \cellcolor{gray!15} & \cellcolor{gray!15} & $\pm{49}$ & 3 \\
\hline
\end{tabular}%
}
\caption{Sensor characteristics and fundamental parameters}
\label{tab:sensor-specs}
\end{table}

\section{Evaluation and results}
\label{sec:evaluation}

\begin{table}
\centering
\renewcommand{\arraystretch}{1.3}
\setlength{\tabcolsep}{7pt}
\resizebox{\textwidth}{!}{%
\begin{tabular}{|c|c|c|c|c|c|c|c|c|c|c|c|}
\hline
\textbf{Test} & \textbf{Location} & \textbf{Sensor} & \textbf{Duration (\si{\second})} & \textbf{Carrier Samples} & \textbf{INS Samples} & \textbf{Acc. Threshold (\si{\meter\per\second^2})} & \textbf{Burn-in} & \textbf{Events} & \textbf{Undefined} & \textbf{Spoofing} & \textbf{Non-Spoofing} \\
\hline
Benign 1.a & Open Sky (Kista) & LSM6D & 300 & 5977 & 30237 & 0.5 & 300 & 536 & 206 & 23 & 307 \\ 
Benign 2.a & Open Sky (Kista) & LSM6D & 295 & 5885 & 30180 & 0.5 & 300 & 527 & 232 & 56 & 239 \\
Benign 3.a & Open Sky (Kista) & LSM6D & 53 & 1058 & 5251 & 0.5 & 50 & 94 & 19 & 16 & 59 \\
\hline
Benign 1.b & Open Sky (Kista) & SCHA63T & 300 & 5977 & 30237 & 0.5 & 300 & 536 & 303 & 34 & 199 \\ 
Benign 2.b & Open Sky (Kista) & SCHA63T & 295 & 5885 & 30180 & 0.5 & 300 & 527 & 245 & 24 & 258 \\
Benign 3.b & Open Sky (Kista) & SCHA63T & 53 & 1058 & 5251 & 0.5 & 50 & 94 & 34 & 12 & 48 \\

\hline
\end{tabular}%
}
\caption{Benign case - baseline scenario}
\label{tab:experimental_benign_baseline}
\end{table}

As shown in \cref{tab:sensor-specs}, the performance of the various inertial sensors varies depending on their category. The Allan deviation in \cref{fig:accelerometers_comp,fig:gyro_comp} reveals that the intrinsic quality of the different \gls{imu} contributes to potential improvements only at higher integration intervals. At short integration times, the sensor performance are comparable. Our scheme uses a sampling window for the carrier phase and the \gls{imu} measurements in the order of \SI{1}{\second}: this corresponds to 20 carrier phase samples and 2000 \gls{imu} measurements per event. Due to the very short integration period, the intrinsic instability of the \gls{imu} minimally affects the detector. Practically, this means relatively cheaper broadly available \gls{imu}s are precise enough to etimate the antenna motion over a short integration period. In contrast, the mobile phone \gls{imu}, although in the same category as the low-cost \gls{imu} in our platform, achieves lower performance. This is possibly due to the implementation of the Android API and the fact that the sampling interval is not strictly controlled, as is the case for our platform.

Our method functions appropriately if antenna movement has a high-frequency component that can be measured against the carrier phase. Empirical evaluation shows that a good value for the minimum acceleration that triggers the detection system is \SI{0.5}{\meter\per\second\squared} and the oscillations frequency range is [\SI{1}{\hertz};\SI{5}{\hertz}] to make sure the assumptions in the sampling rate and \gls{pll} bandwidth are respected (\SI{20}{\hertz} for the \gls{gnss} measurements and \SI{100}{\hertz} for the \gls{imu}s). 
Although at a lower sampling rate compared to specialized custom designs as in \cite{WOS:000375213003001}, the measured carrier phase is continuous and available for all satellites in view, even if not used in the internal \gls{pnt} engine. 

\subsection{Benign case - baseline scenario}

\begin{figure}
    \centering
    \begin{subfigure}[t]{.45\linewidth}
        \centering
        \includegraphics[width=\linewidth]{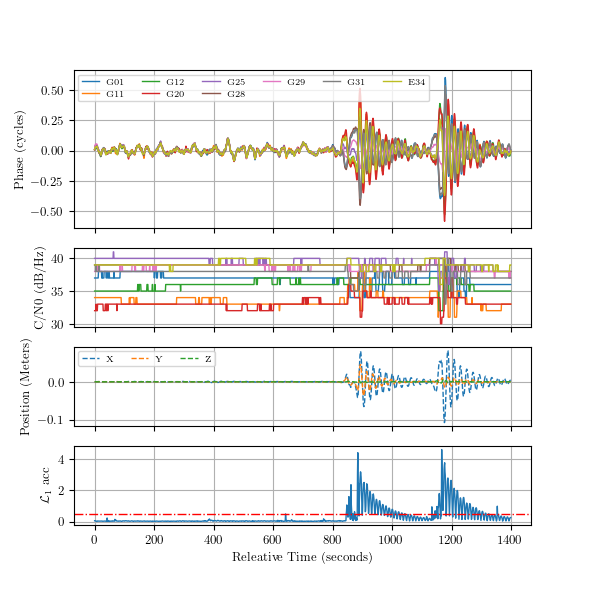}
        \caption{Test Benign 1.a, \cref{tab:experimental_spoofing_baseline}.}
        \label{fig:data_clean_1_kista_lis}
    \end{subfigure}
        \begin{subfigure}[t]{.45\linewidth}
        \centering
        \includegraphics[width=\linewidth]{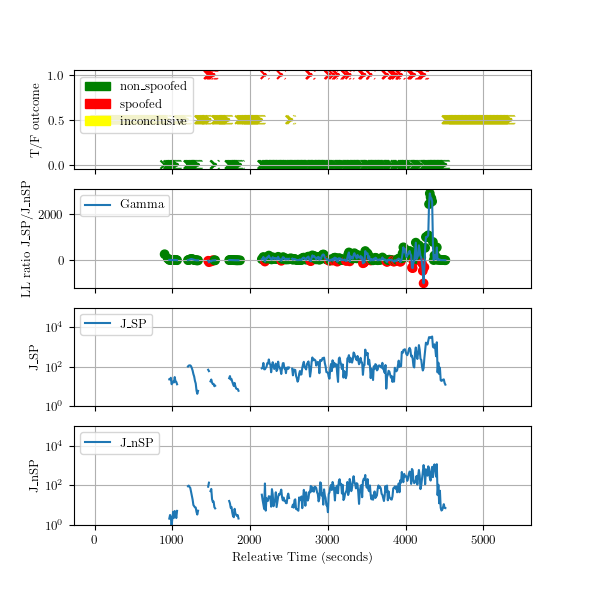}
        \caption{Test Benign 1.a, \cref{tab:experimental_spoofing_baseline}.}
        \label{fig:outcome_clean_1_kista_lis}
    \end{subfigure}\\
    \begin{subfigure}[t]{.45\linewidth}
        \centering
        \includegraphics[width=\linewidth]{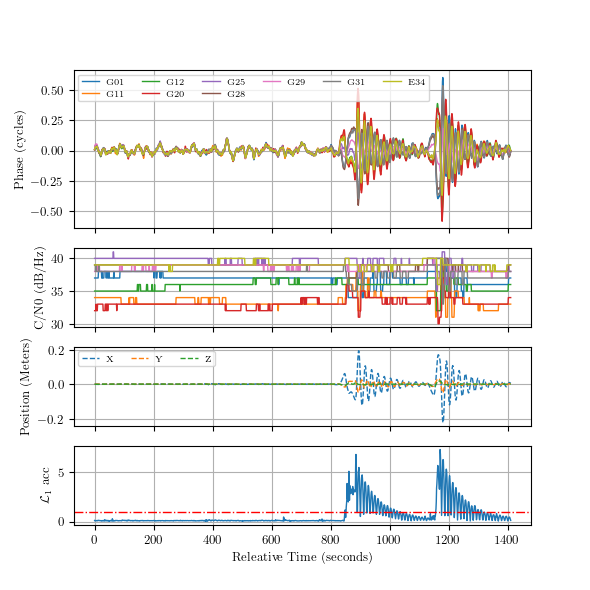}
        \caption{Test Benign 1.b, \cref{tab:experimental_spoofing_baseline}.}
        \label{fig:data_clean_1_kista_murata}
    \end{subfigure}
    \begin{subfigure}[t]{.45\linewidth}
        \centering
        \includegraphics[width=\linewidth]{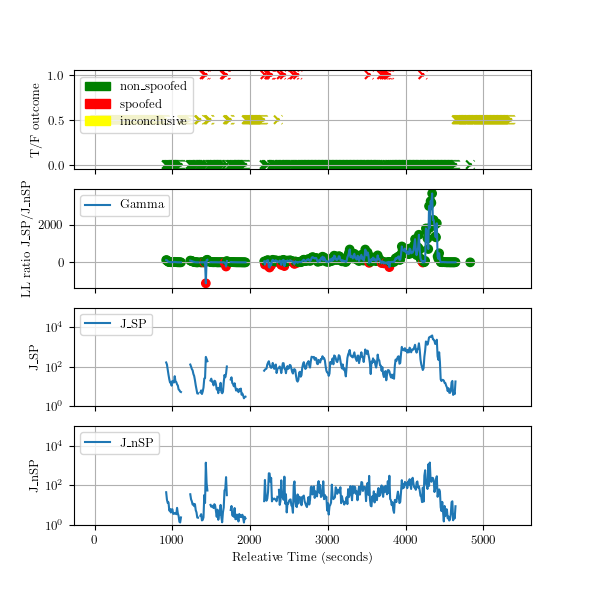}
        \caption{Test Benign 1.b, \cref{tab:experimental_spoofing_baseline}.}
        \label{fig:outcome_clean_1_kista_murata}
    \end{subfigure}
    \caption{Partial sequence of Carrier - INS coupled measurements (left 1,2) and detection performance (right 3,4) in a known good scenario.}
\end{figure}

\cref{fig:data_clean_1_kista_lis,fig:data_clean_1_kista_murata,} show a sample of the results for the detection system based on LSM6D and SCHA63T \gls{imu}, respectively. The system is operating in a known-good environment, without the presence of any adversary. The minimum 3D vector acceleration selected during testing is \SI{0.5}{\meter\per\second^2} and if the acceleration is not above the required threshold, the outcome of the spoofing detection system is not determined. In particular, the inconclusive cases are not included in the accuracy evaluation, as there is no detection performed. 
In the benign scenario, the detector correctly identifies the absence of a spoofer with high confidence depending on the test case (in case of the SCHA63T \gls{imu}). When the carrier-\gls{imu} coupling is based on the SCHA63T sensor, the achievable true positive detection rate for the benign case is 96\% in the best case scenario (73\% worst case). The LSM6D based measurements provide similar quality, although about 8\% worse performance for the benign case, with the detector more skewed towards the spoofed hypothesis. Overall, the performance is consistent over the presented test cases in \cref{tab:experimental_benign_baseline}. The result for both sensors is remarkably similar, showing that the \gls{imu} performance does not dominate the accuracy of the detector at such short integration times.



\subsection{Fully adversarial case - baseline scenario}

\begin{table*}
\centering
\renewcommand{\arraystretch}{1.3}
\setlength{\tabcolsep}{7pt}
\resizebox{\textwidth}{!}{%
\begin{tabular}{|c|c|c|c|c|c|c|c|c|c|c|c|}
\hline
\textbf{Test} & \textbf{Location} & \textbf{Sensor} & \textbf{Duration (\si{\second})} & \textbf{Carrier Samples} & \textbf{INS Samples} & \textbf{Acc. Threshold (\si{\meter\per\second^2})} & \textbf{Burn-in} & \textbf{Events} & \textbf{Undefined} & \textbf{Spoofing} & \textbf{Non-Spoofing} \\
\hline
Spoof 1.a & NSS Lab (Kista) & LSM6D & 310 & 6200 & 30990 & 0.5 & 50 & 608 & 261 & 180 & 80 \\ 
Spoof 2.a & NSS Lab (Kista) & LSM6D & 248 & 5779 & 29730 & 1.0 & 800 & 416 & 367 & 39 & 10 \\
Spoof 3.a & NSS Lab (Kista) & LSM6D & 88 & 1767 & 12224 & 1.0 & 10 & 173 & 54 & 98 & 21 \\
\hline
Spoof 1.b & NSS Lab (Kista) & SCHA63T & 310 & 6200 & 30990 & 0.5 & 50 & 608 & 392 & 132 & 84 \\
Spoof 2.b & NSS Lab (Kista) & SCHA63T & 248 & 5779 & 29730 & 1.0 & 800 & 416 & 390 & 22 & 4 \\
Spoof 3.b & NSS Lab (Kista) & SCHA63T & 88 & 1767 & 12224 & 1.0 & 60 & 164 & 105 & 37 & 22 \\
\hline

\end{tabular}%
}
\caption{Fully adversarial case - baseline scenario}
\label{tab:experimental_spoofing_baseline}
\end{table*}

\begin{figure}
\centering
    \begin{subfigure}[t]{.45\linewidth}
        \centering
        \includegraphics[width=\linewidth]{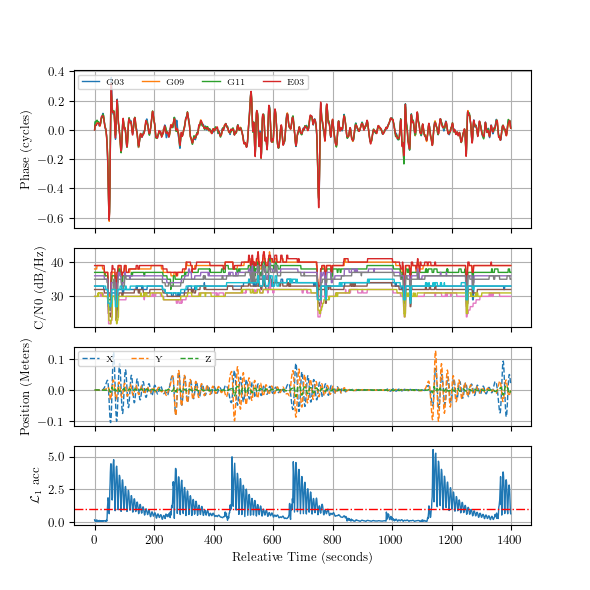}
        \caption{Test Spoof 3.a, \cref{tab:experimental_spoofing_baseline}.}
        \label{fig:spoof_test_kista_1_lis}
    \end{subfigure}
    \begin{subfigure}[t]{.45\linewidth}
        \centering
        \includegraphics[width=\linewidth]{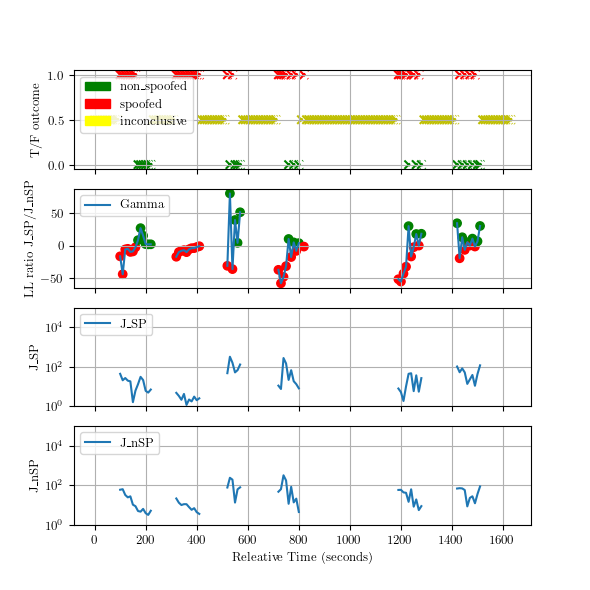}
        \caption{Test Spoof 3.a, \cref{tab:experimental_spoofing_baseline}.}
        \label{fig:outcome_spoof_kista_lis}
    \end{subfigure}\\
    \begin{subfigure}[t]{.45\linewidth}
        \centering
        \includegraphics[width=\linewidth]{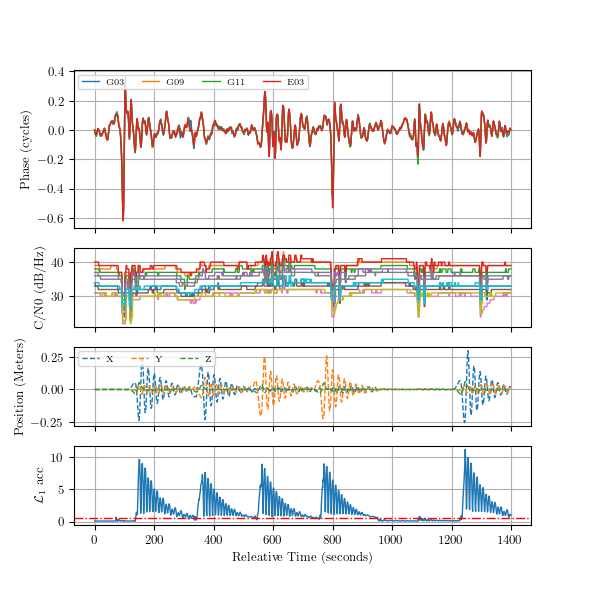}
        \caption{Test Spoof 3.b, \cref{tab:experimental_spoofing_baseline}.}
        \label{fig:spoof_test_kista_1_murata}
    \end{subfigure}
    \begin{subfigure}[t]{.45\linewidth}
        \centering
        \includegraphics[width=\linewidth]{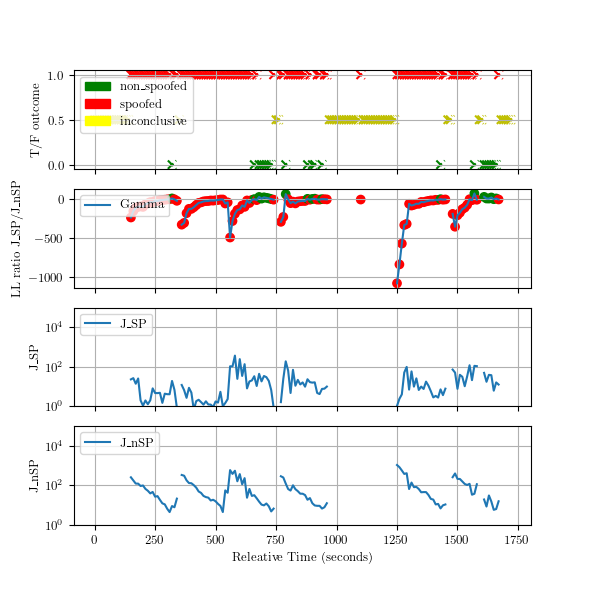}
        \caption{Test Spoof 3.b, \cref{tab:experimental_spoofing_baseline}.}
        \label{fig:outcome_spoof_1_kista_murata}
    \end{subfigure}
    \caption{Detection performance in a known spoofed scenario (left 1,2) and detection performance (right 3,4)}
\end{figure}

In the second validation test set, a spoofed constellation is transmitted to the victim receiver. 
The exact objective of the attacker during the spoofing phase is not strictly important for the validity of the results, but the adversary spoofs the receiver forcing a location near the legitimate one, with coarse alignment of the time solution (e.g., without proper code-phase alignment), meaning that the receiver \gls{pnt}-based time is correct within the current frame. Generally, this is not need, but it simplifies the handling of the RINEX files so that the measurements \gls{imu} measurements are still aligned with the \gls{gnss} carrier timestamps. An extract of the carrier phase measurements and inertial estimation under spoofing conditions is shown in \cref{fig:spoof_test_kista_1_lis,fig:spoof_test_kista_1_murata}, where the difference in the carrier phase structure mentioned in \cref{sec:methodology} is visible when compared to \cref{fig:data_clean_1_kista_lis,fig:data_clean_1_kista_murata}. 
In the spoofed case, the detector performs well, with a true positive rate up to 90\% in spoofing detection for the SCHA63T sensor. Similarly to the benign case, the LSM6D sensor performs worse here too, but with an higher reduction in accuracy (about 15\% less accurate). A summary of the spoofing baseline scenarios is provided in \cref{tab:experimental_spoofing_baseline}. 
Both sensors under test show equivalent performances when compared with an industrial grade high quality inertial platform (\cref{tab:sensor-specs}, the MTI-3 inertial unit). These results are not detailed for brevity.

Despite the difference in quality of the different \gls{imu} tested, the outcome is strikingly similar. Given the short integration window, there are only limited benefits due to much more stable \gls{imu}.
The variance in performance is possibly due to the different accuracy and stability of the gyroscope in the three platforms. As the platform needs to convert the measured acceleration into linear acceleration in the sensor frame, the gyroscope is used to estimate the orientation of the sensor platform body in space, so that gravity can be subtracted from the acceleration measurements. This leads to a variability in the estimation in the linear accelerations that overall influences the accuracy of the detector. Such observation is supported by the analysis of the Allan deviation in \cref{fig:gyro_comp}, highlighting different performances in the gyroscope sensor. 

\subsection{Live testing at Jammertest}
\begin{figure*}
\centering
    \begin{subfigure}[t]{.45\linewidth}
        \centering
        \includegraphics[width=\linewidth]{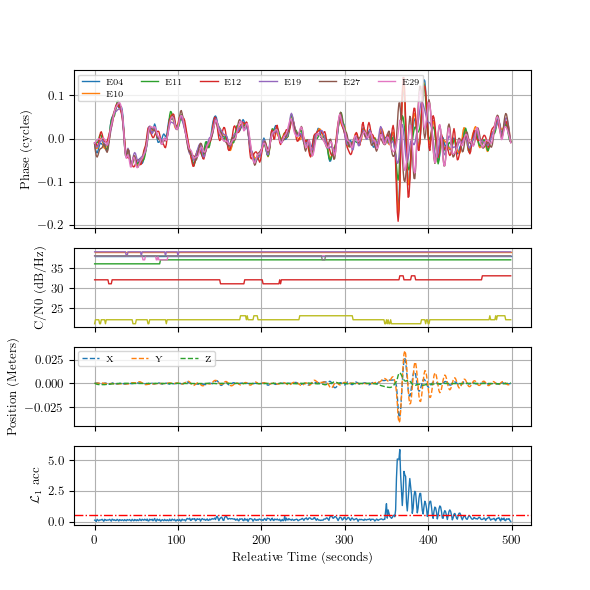}
        \caption{Data sample of receiver under attack, with static platform and high frequency oscillations.}
        \label{fig:jammertest_4_data_slice}
    \end{subfigure}
        \begin{subfigure}[t]{.45\linewidth}
        \centering
        \includegraphics[width=\linewidth]{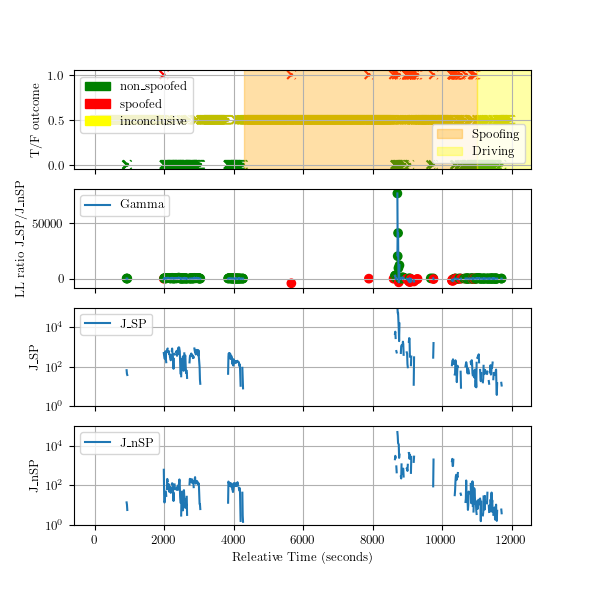}
        \caption{Detection outcome, with marked beginning of the attack. The detector scans the entire trace in Test 1 from \cref{tab:jammertest_testing}.}
        \label{fig:jammertest_4_outcome}
    \end{subfigure}
    \caption{Live Test 1 from \cref{tab:jammertest_testing}, with different benign and spoofed conditions. The data sample shows on one higher resolution slice of data during the spoofing attack.}
\end{figure*}

Tests conducted in Jammertest 2024 evaluate the real-life performance of this method. A summary of the test conditions is given in \cref{tab:jammertest_testing}. The table also reports the test identification number for the official test description \cite{jammertest_test_catalog}. The performed tests include both meaconing and spoofing in both dynamic and static setting. 
Test 1 in \cref{tab:jammertest_testing} includes three separate moments. In the first part of the test, the device is static and in a benign scenario, the detector is measuring only legitimate carrier phase data. At the start of the attack, the carrier phase information is corrupted by the spoofing signals, and after the tracking loops are captured, the receiver is spoofed as seen in the slice shown in \cref{fig:jammertest_4_data_slice}.
The detector starts flagging spoofing events in the measurements, as shown in \cref{fig:jammertest_4_outcome}. Additionally, this can also be seen in a sharp change in the $C/N_0$ for the satellites in view, but while intuitively the  $C/N_0$ should improve, due to countermeasures implemented internally in the receiver, the $C/N_0$ drops reflecting the change in the front end programmable gain amplifier within the receiver. The attack period spans for about \SI{2}{\minute}, after which the vehicle moves away from the adversarial zone and the platform produces again a valid solution, shown in the third part of \cref{fig:jammertest_4_outcome}. 

\begin{figure}
    \centering
    \begin{subfigure}[t]{0.3\linewidth}
        \centering
        \includegraphics[width=\linewidth]{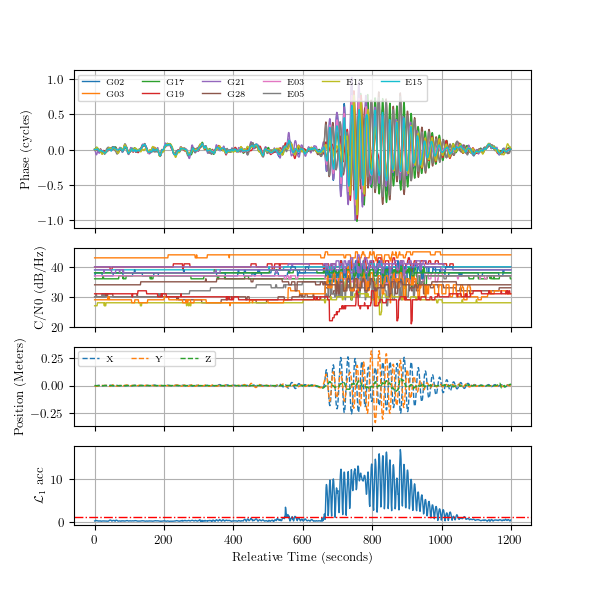}
        \caption{Data sample of the benign part of the OTA test, with static platform and high frequency oscillations from Test 2 from \cref{tab:jammertest_testing}}
        \label{fig:jammertest_8_data_slice_clean}
    \end{subfigure}
    \begin{subfigure}[t]{0.3\linewidth}
        \centering
        \includegraphics[width=\linewidth]{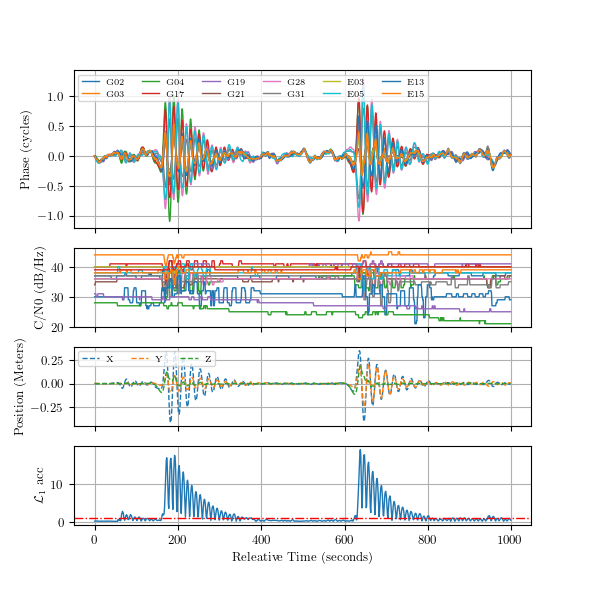}
        \caption{Data sample of the platform under spoofing, with high frequency oscillations at the start of Test 2 from \cref{tab:jammertest_testing}}
        \label{fig:jammertest_8_data_slice_clean_spoofed}
    \end{subfigure}
    \begin{subfigure}[t]{0.3\linewidth}
        \centering
        \includegraphics[width=\linewidth]{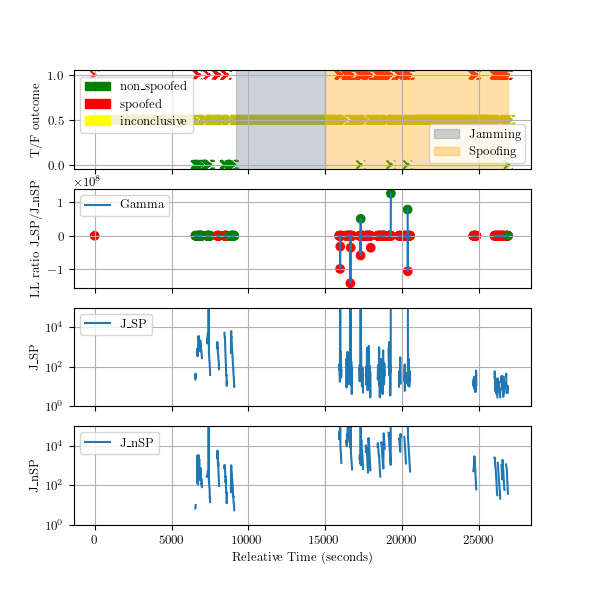}
        \caption{Detection outcome, with marked beginning of the attack. The detector scans the entire trace in Test 2 from \cref{tab:jammertest_testing}}
        \label{fig:jammertest_8_outcome}
    \end{subfigure}
    \caption{Jammertest evaluation under mixed spoofing and non spoofing setups}
\end{figure}

During spoofing, the method performs as expected. The attack is correctly detected similarly to the validation test cases. \cref{fig:jammertest_8_data_slice_clean,fig:jammertest_8_data_slice_clean_spoofed} show carrier-\gls{imu} measurements for two selected moments of Test 2 in \cref{tab:jammertest_testing}, without and with the presence of a spoofer respectively. With the receiver operating in a benign scenario for \SI{5}{\minute}, the attacker first forces a loss of lock by jamming (about \SI{5}{\minute}), during which raw measurements are not available. After this, the adversary begins transmission of the spoofing signals, with coherent alignment to the legitimate constellation. Observations show that while the receiver does not provide a solution, the spoofing signals are still acquired and tracked. The attack mounted is quite subtle, as it forces a progressive change in the pseudoranges, effectively forcing a clock drift. Such attack is successful against a wide range of receivers as the adversarial signals are largely similar to the legitimate ones. Once the receiver lock on the spoofing signals, the outcome of the attack detection is shown in \cref{fig:jammertest_8_outcome}, showing successful detection of the attack. 

Compared to the baseline scenarios, testing in static setting leads to similar results but overall the accuracy of the method is lower. Even if the adversary cannot guarantee the correct spreading of the carrier phase, the propagation environment actually makes the adversarial task simpler: reflections and differences in propagation result in propagation channels exactly identical for all the satellites. This leads to a higher number of false negatives, as the detector tends to be biased toward the null hypothesis. In particular, for mobile spoofing the task of the adversary is made more complex by environmental shadowing that masks the adversarial \gls{los} allowing the receiver to briefly re-acquire the legitimate signals. Additionally, the dynamic setting caused several issues in the phase solution consistency. Even if enough movement was available at the antenna, evaluations of the coupled ins-carrier measurements within the tests scope are inconclusive. This effect is likely due to cycle slips in the carrier phase, possibly because of the vehicle's mobility. 

Furthermore, the number of events in known-benign and known-spoofed conditions are different. For this reason, the tests from \cref{tab:jammertest_testing} aim at showing that the detector is capable of distinguishing the transition between the spoofed and non spoofed case, as a method that can complement other \gls{pnt} monitoring methods. For each test, the start of the spoofing event is marked at the relevant time. \cref{fig:jammertest_4_outcome} shows that after the initial re-acquisition at the \gls{gnss} receiver the attack is correctly detected by the platform for all the events where the movement is available. When the vehicle starts moving and the victim antenna is out of range of the spoofer, the detector successfully transitions back to the null hypothesis.

\begin{table*}[h!]
\renewcommand{\arraystretch}{1.3}
\setlength{\tabcolsep}{7pt}
\resizebox{\textwidth}{!}{%
\begin{tabular}{|c|c|p{0.4\textwidth}|c|c|c|c|c|c|c|c|c|}
\hline
\textbf{Test} & \textbf{Location} & \textbf{Scenario} & \textbf{Duration (\si{\second})} & \textbf{Carrier Samples} & \textbf{INS Samples} & \textbf{Acc. Threshold (\si{\meter\per\second^2})} & \textbf{Burn-in} & \textbf{Events} & \textbf{Undefined} & \textbf{Spoofing} & \textbf{Non-Spoofing} \\
\hline
Test 1 & Bleik & Simulated driving - Initial Jamming, Galileo only (2.3.8) & 939 & 18799 & 96184 & 0.5 & 0 & 1198 & 879 & 55 & 264 \\
\hline
Test 2 & Bleik & PR error - Initial Jamming and forced clock drift (2.4.13) & 1353 & 27073 & 162315 & 0.5 & 0 & 2706 & 2287 & 327 & 92 \\
\hline
\end{tabular}%
}
\caption{Jammertest OTA spoofing tests}
\label{tab:jammertest_testing}
\end{table*}

Overall, the detection system is capable of detecting adversarial signal manipulation even in real-life conditions, where multipath and other signal imperfections are present. Compared to the static test scenarios, the mobility tests are less accurate. This could be due to multiple factors, but this is likely due to loss of carrier phase information which can happen in moving receivers: this is a limitation of the current implementation, where we require continuous carrier phase information for the segments the detector operates on. Additionally, a better separation between the acceleration due to the movement of the car and linear acceleration due to the high-frequency motion of the platform is beneficial in increasing the accuracy. 

\subsection{Mobile phone platforms} 

Unfortunately, tests conducted with the Pixel platform show that the currently available measurement capabilities of the Android API are not sufficient to provide high-quality carrier phase data. 
The onboard sensors provide a high sampling rate, the \gls{gnss} raw message information is too sparse for the detection system presented here to be effective. As the detector is based on high-frequency oscillations, the \SI{1}{\hertz} sampling rate provided by the Pixel smartphone does not give sufficient temporal resolution. On the other hand, the inertial sensors already provide high enough sampling rate to support the method presented here. The deep integration of low rate carrier based position and inertial sensors has been explored in mobile phones and has promising results for precision navigation \cite{bochkati2020demonstration}, but still the possibility of faster measurement rate is lacking at the \gls{gnss} chipset. While this is possibly fixed by the chipset's GNSS receiver, to the best of our knowledge there is no support for higher update rate. Availability of such a feature would make the implementation of the method presented here possible even on mobile phones, truly expanding the possibilities for robust navigation in everyday's systems.

\section{Conclusions}
\label{sec:conclusion}

We presented a method that couples carrier phase measurements with an inexpensive \gls{imu} to allow a commercial platform to detect spoofing and recover from adversarial manipulation. Out method is agnostic of the actual \gls{pnt} solution and it can be used without knowledge of the precise internal operation of the receiver. Additionally, we show that low-cost mass market \gls{imu}s that are traditionally not suited for navigation purposes can be used reliably by our method to detect spoofing at a minimal loss of accuracy. Nevertheless, there are limitations. The separation of the high frequency movement from the actual accelerations due to the car movement itself require further exploration. The current method does not distinguish between the nature of the movement of the antenna (between 1D or 3D) which would allow for significant performance improvements. Additionally, the lack of support for multi-rate raw measurements sampling in the Android API is the only limitation to make suck work applicable to truly mobile devices. In conclusion, the presented method represents a considerable step forward towards reliable and assured \gls{pnt} in mobile devices, providing a simple yet effective detection of adversarial signals.

\section*{Acknowledgements}
The NSS group is part of Safran Minerva Academic program and the Skydel software was granted through the academic partnership, including the additional plugin licenses that made this work possible. This work was supported in parts by the national strategic research area on security and emergency preparedness.

\bibliographystyle{apalike}
\bibliography{references_old}

\end{document}

%% file: acronyms.tex
\newacronym{scd}{SCD}{Spectral Correlation Density}
\newacronym{cw}{CW}{Continuous Wave}
\newacronym{fam}{FAM}{Fourier Accumulation Method}
\newacronym{snr}{SNR}{Signal to Noise Ratio}
\newacronym{bpsk}{BPSK}{Binary Phase Shift Key}

\newacronym{ppd}{PPD}{Privacy Protection Device}

\newacronym{gnss}{GNSS}{Global Navigation Satellite System}
\newacronym{gps}{GPS}{Global Positioning System}
\newacronym{nmea}{NMEA}{National Marine Electronics Association}
\newacronym{nasa}{NASA}{National Aeronautics and Space Administration}
\newacronym{agps}{A-GPS}{Assisted/Augmented GPS}
\newacronym{scer}{SCER}{Security Code Estimation and Replay}
\newacronym{nma}{OS-NMA}{Open Service Navigation Message Authentication}
\newacronym{sis}{SIS}{Signal In Space}
\newacronym{raim}{RAIM}{Receiver Autonomous Integrity Monitoring}
\newacronym{pnt}{PNT}{Position-Navigation-Time}
\newacronym{rtk}{RTK}{Real-time Kinematics}
\newacronym{agc}{AGC}{Automatic Gain Control}
\newacronym{meo}{MEO}{Medium Earth Orbit}
\newacronym{rssi}{RSSI}{Received Signal Strength Indicator}
\newacronym{llh}{LLH}{Latitude, Longitude and Height}
\newacronym{osnma}{OSNMA}{Open Signal Navigation Message Authentication}
\newacronym{sqm}{SQM}{Signal Quality Monitoring}
\newacronym{los}{LOS}{Line of Sight}

\newacronym{nco}{NCO}{Numerically Controlled Oscillator}
\newacronym{pll}{PLL}{Phase-Locked Loop}
\newacronym{ecef}{ECEF}{Earth Centered Earth Fixed}
\newacronym{pca}{PCA}{Principal Components Analysis}
\newacronym{mle}{MLE}{Maximum Likelyhood Estimator}

\newacronym{sdr}{SDR}{Software Defined Radio}
\newacronym{sdrs}{SDR}{Software Defined Radios}
\newacronym{ots}{OTS}{Off-the-Shelf}
\newacronym{rt}{RT}{Real-Time}
\newacronym{ide}{IDE}{Integrated Development Environment}
\newacronym{fifo}{FIFO}{First-In-First-Out}
\newacronym{sma}{SMA}{SubMiniature Version A}
\newacronym{rf}{RF}{Radio Frequency}
\newacronym{uav}{UAV}{Unmanned Aerial Vehicle}
\newacronym{vtol}{VTOL}{Vertical Take Off and Landing}
\newacronym{ota}{OTA}{Over-The-Air}
\newacronym{ism}{ISM}{Industrial Scientific and Medical}

\newacronym{swap}{SWaP}{Size, Weight and Power}
\newacronym{imu}{IMU}{Inertial Measurement Unit}
\newacronym{cots}{COTS}{Commercial Off the Shelf}
\newacronym{som}{SoM}{System on Module}

%% file: IONconf_template.bbl
\begin{thebibliography}{}

\bibitem[Akos, 2012]{WOS:000209006500003}
Akos, D.~M. (2012).
\newblock {Who's Afraid of the Spoofer? GPS/GNSS Spoofing Detection via
  Automatic Gain Control (AGC)}.
\newblock {\em Journal Of The Institute Of Navigation}, 59(4):281--290.

\bibitem[Ali et~al., 2014]{WOS:000359380700146}
Ali, K., Manfredini, E.~G., and Dovis, F. (2014).
\newblock {Vestigial Signal Defense through Signal Quality Monitoring
  Techniques based on Joint Use of Two Metrics}.
\newblock In {\em IEEE/ION Position, Location And Navigation Symposium - Plans
  2014}.

\bibitem[Allan, 1987]{Allan1987}
Allan, D.~W. (1987).
\newblock Time and frequency (time-domain) characterization, estimation, and
  prediction of precision clocks and oscillators.
\newblock {\em IEEE Transactions on Ultrasonics, Ferroelectrics, and Frequency
  Control}, 34(6):647--654.

\bibitem[Amin et~al., 2016]{7471534}
Amin, M.~G., Closas, P., Broumandan, A., and Volakis, J.~L. (2016).
\newblock Vulnerabilities, threats, and authentication in satellite-based
  navigation systems [scanning the issue].
\newblock {\em Proceedings of the IEEE}, 104(6):1169--1173.

\bibitem[Anderson et~al., 2017]{WOS:000419292302031}
Anderson, J.~M., Carroll, C. K.~L., DeVilbiss, N.~P., Gillis, J.~T., Hinks,
  J.~C., O'Hanlon, B.~W., Rushanan, J.~J., Scott, L., and Yazdi, R.~A. (2017).
\newblock Chips-message robust authentication (chimera) for gps civilian
  signals.
\newblock In {\em 30th International Technical meeting of the Satellite
  Division of the Insititute of Navigation (ION GNSS+)}.

\bibitem[Bastide et~al., 2003]{akos2003}
Bastide, F., Akos, D., Macabiau, C., and Roturier, B. (2003).
\newblock {Automatic Gain Control (AGC) as an Interference Assessment Tool}.
\newblock In {\em 16th International Technical Meeting of the Satellite
  Division of The Institute of Navigation (ION GPS/GNSS}, pages 2042--2053,
  Portland, OR, USA. Institute of Navigation.

\bibitem[Bochkati et~al., 2020]{bochkati2020demonstration}
Bochkati, M., Sharma, H., Lichtenberger, C.~A., and Pany, T. (2020).
\newblock Demonstration of fused rtk (fixed) + inertial positioning using
  android smartphone sensors only.
\newblock In {\em IEEE/ION Position, Location And Navigation Symposium, Plans},
  pages 1140--1154, Portland, OR, USA.

\bibitem[Clements et~al., 2022]{Clements2022CarrierphaseAI}
Clements, Z., Yoder, J.~E., and Humphreys, T.~E. (2022).
\newblock Carrier-phase and imu based gnss spoofing detection for ground
  vehicles.
\newblock {\em The International Technical Meeting of the The Institute of
  Navigation}.

\bibitem[Cucchi et~al., 2021]{Cucchi2021AssessingReceiver}
Cucchi, L., Damy, S., Paonni, M., et~al. (2021).
\newblock Assessing galileo osnma under different user environments by means of
  a multi-purpose test bench, including a software-defined {GNSS} receiver.
\newblock In {\em 34th International Technical Meeting of the Satellite
  Division of the Institute of Navigation, (ION {GNSS}+)}.

\bibitem[Curran and Broumandan, 2017]{Curran2017OnTU}
Curran, J.~T. and Broumandan, A. (2017).
\newblock On the use of low-cost imus for {GNSS} spoofing detection in
  vehicular applications.
\newblock In {\em International Technical Symposium on Navigation and Timing
  (ITSNT)}.

\bibitem[Ferraris et~al., 1994]{Ferraris1994CalibrationStandards}
Ferraris, F., Gorini, I., Grimaldi, U., and Parvis, M. (1994).
\newblock {Calibration of three-axial rate gyros without angular velocity
  standards}.
\newblock {\em Sensors and Actuators A: Physical}, 42(1-3):446--449.

\bibitem[G{\"o}tzelmann et~al., 2023]{Gotzelmannnavi.572}
G{\"o}tzelmann, M., K{\"o}ller, E., Viciano-Semper, I., Oskam, D., Gkougkas,
  E., and Simon, J. (2023).
\newblock Galileo open service navigation message authentication: Preparation
  phase and drivers for future service provision.
\newblock {\em Journal of the Institute of Navigation}, 70(3).

\bibitem[Hernández et~al., 2019]{8714151}
Hernández, I.~F., Ashur, T., Rijmen, V., Sarto, C., Cancela, S., and Calle, D.
  (2019).
\newblock Toward an operational navigation message authentication service:
  Proposal and justification of additional osnma protocol features.
\newblock In {\em European Navigation Conference (ENC)}.

\bibitem[Hu et~al., 2018a]{WOS:000423143700015}
Hu, Y., Bian, S., Cao, K., and Ji, B. (2018a).
\newblock {GNSS spoofing detection based on new signal quality assessment
  model}.
\newblock {\em Gps Solutions}, 22(1).

\bibitem[Hu et~al., 2018b]{Hu_Bian_Ji_Li_2018}
Hu, Y., Bian, S., Ji, B., and Li, J. (2018b).
\newblock Gnss spoofing detection technique using fraction parts of
  double-difference carrier phases.
\newblock {\em Journal of Navigation}, 71(5):1111–1129.

\bibitem[Huang and Yang, 2015]{HuangL2015}
Huang, L. and Yang, Q. (2015).
\newblock {Low-cost GPS simulator - GPS spoofing by SDR}.
\newblock In {\em Proceedings of DEF CON23}, Las Vegas, NV , USA.

\bibitem[Humphreys et~al., 2012]{Humphreys2012}
Humphreys, T., Bhatti, J., Shepard, D., et~al. (2012).
\newblock {The Texas spoofing test battery: Toward a standard for evaluating
  GPS signal authentication techniques}.
\newblock In {\em 25th International Technical Meeting of the Satellite
  Division of the Institute of Navigation 2012, (ION GNSS+)}, Nashville, TN,
  USA.

\bibitem[Humphreys et~al., 2008]{HumphreysAssessingSpoofer}
Humphreys, T.~E., Ledvina, B.~M., Psiaki, M.~L., et~al. (2008).
\newblock {Assessing the Spoofing Threat: Development of a Portable GPS
  Civilian Spoofer}.
\newblock In {\em 21st International Technical Meeting of the Satellite
  Division of The Institute of Navigation (ION GNSS)}, Savannah, GE, USA.

\bibitem[Jovanovic et~al., 2014]{Jovanovic2014MultitestDA}
Jovanovic, A., Botteron, C., and Farine, P.-A. (2014).
\newblock Multi-test detection and protection algorithm against spoofing
  attacks on gnss receivers.
\newblock {\em 2014 IEEE/ION Position, Location and Navigation Symposium
  (IEEE/ION PLANS)}, pages 1258--1271.

\bibitem[Kujur et~al., 2024]{Kujurnavi.629}
Kujur, B., Khanafseh, S., and Pervan, B. (2024).
\newblock Optimal ins monitor for gnss spoofer tracking error detection.
\newblock {\em NAVIGATION: Journal of the Institute of Navigation}, 71(1).

\bibitem[Lee et~al., 2022]{Lee2022GNSSFD}
Lee, D.-K., Taylor, T., Akos, D.~M., Yun, J., Jo, Y., and Park, B.-K. (2022).
\newblock Gnss fault detection and mitigation using android imu.
\newblock {\em The International Technical Meeting of the Satellite Division of
  The Institute of Navigation (ION GNSS+)}.

\bibitem[Lenhart et~al., 2022]{LenhartSP:C:2022}
Lenhart, M., Spanghero, M., and Papadimitratos, P. (2022).
\newblock {Distributed and Mobile Message Level Relaying/Replaying of GNSS
  Signals}.
\newblock In {\em International Technical Meeting of The Institute of
  Navigation (ION ITM)}, pages 56--57, Long Beach, CA, USA.

\bibitem[Madgwick, 2010]{Madgwick2010AnEO}
Madgwick, S. O.~H. (2010).
\newblock An efficient orientation filter for inertial and inertial / magnetic
  sensor arrays.
\newblock Technical report.

\bibitem[Meurer and Antreich, 2017]{Meurer2017}
Meurer, M. and Antreich, F. (2017).
\newblock {\em Signals and Modulation}, page 91–119.
\newblock Springer International Publishing.

\bibitem[Mina et~al., 2021]{WOS:000874785704023}
Mina, T., Kanhere, A., Kousik, S., and Gao, G. (2021).
\newblock Continuous gps authentication with chimera using stochastic
  reachability analysis.
\newblock In {\em 34th International Technical meeting of the Satellite
  Division of the Insititute of Navigation (ION GNSS+)}.

\bibitem[Miralles et~al., 2018]{WOS:000545000100015}
Miralles, D., Levigne, N., Akos, D.~M., Blanch, J., and Lo, S. (2018).
\newblock Android raw gnss measurements as a new anti-spoofing and anti-jamming
  solution.
\newblock In {\em International technical meeting of the satellite division of
  the Instituite of Navigation (ION GNSS+)}, Institute of Navigation Satellite
  Division Proceedings of the International Technical Meeting, pages 334--344.

\bibitem[Motallebighomi et~al., 2023]{10.1145/3558482.3590186}
Motallebighomi, M., Sathaye, H., Singh, M., and Ranganathan, A. (2023).
\newblock {Location-independent GNSS Relay Attacks: A Lazy Attacker's Guide to
  Bypassing Navigation Message Authentication}.
\newblock In {\em ACM Conference on Security and Privacy in Wireless and Mobile
  Networks (WISEC)}, WiSec '23, New York, NY, USA. Association for Computing
  Machinery.

\bibitem[O'Driscoll et~al., 2023]{10139953}
O'Driscoll, C., Winkel, J., and Hernandez, I.~F. (2023).
\newblock Assisted nma proof of concept on android smartphones.
\newblock In {\em 2023 IEEE/ION Position, Location and Navigation Symposium
  (PLANS)}, pages 559--569, Monterey, CA, USA.

\bibitem[Papadimitratos and Jovanovic, 2008a]{PapadimitratosJ:C:2008}
Papadimitratos, P. and Jovanovic, A. (2008a).
\newblock {GNSS-based Positioning: Attacks and Countermeasures}.
\newblock In {\em IEEE Military Communications Conference (IEEE MILCOM)}, San
  Diego, CA, USA.

\bibitem[Papadimitratos and Jovanovic, 2008b]{PapadimitratosJa:C:2008}
Papadimitratos, P. and Jovanovic, A. (2008b).
\newblock {Protection and Fundamental Vulnerability of GNSS}.
\newblock In {\em IEEE International Workshop on Satellite and Space
  Communications (IEEE IWSSC)}, Toulouse, France.

\bibitem[Peng et~al., 2019]{Jiadong2019}
Peng, C., Li, H., Wen, J., and Lu, M. (2019).
\newblock {Research of Intermediate Spoofing Without Precise Target
  Information}.
\newblock In {\em China Satellite Navigation Conference (CSNC)}. Springer
  Singapore.

\bibitem[Psiaki et~al., 2014]{WOS:000356331204003}
Psiaki, M.~L., O'Hanlon, B.~W., Powell, S.~P., Bhatti, J.~A., Wesson, K.~D.,
  Humphreys, T.~E., and Schofield, A. (2014).
\newblock Gnss spoofing detection using two-antenna differential carrier phase.
\newblock In {\em International Technical meeting of the Satellite Division of
  the Insititute of Navigation (ION GNSS)}, pages 2776--2800.

\bibitem[Psiaki et~al., 2013]{WOS:000375213003001}
Psiaki, M.~L., Powell, S.~P., and O'Hanlon, B.~W. (2013).
\newblock Gnss spoofing detection using high-frequency antenna motion and
  carrier-phase data.
\newblock In {\em International Technical Meeting of the Satellite Division of
  The Institute of Navigation (ION GNSS+)}, Nashville, TN, USA.

\bibitem[Rustamov et~al., 2023]{10081330}
Rustamov, A., Minetto, A., and Dovis, F. (2023).
\newblock Improving gnss spoofing awareness in smartphones via statistical
  processing of raw measurements.
\newblock {\em IEEE Open Journal of the Communications Society}, 4:873--891.

\bibitem[Sanz~Subirana et~al., 2011]{navipedia_carrier_fixing}
Sanz~Subirana, J., Juan~Zornoza, J., and Hernández, M. (2011).
\newblock Carrier phase ambiguity fixing.
\newblock
  \url{https://gssc.esa.int/navipedia/index.php?title=Carrier_Phase_Ambiguity_Fixing}.

\bibitem[Sathaye et~al., 2022]{DBLP:conf/ndss/SathayeLCR22}
Sathaye, H., LaMountain, G., Closas, P., and Ranganathan, A. (2022).
\newblock Semperfi: Anti-spoofing {GPS} receiver for uavs.
\newblock In {\em 29th Annual Network and Distributed System Security Symposium
  (NDSS)}.

\bibitem[Sharma et~al., 2021]{Sharma2021TimeSynchronizedGD}
Sharma, H., Bochkati, M., and Pany, T. (2021).
\newblock Time-synchronized gnss/imu data logging from android smartphone and
  its influence on the positioning accuracy.
\newblock {\em International Technical Meeting of the Satellite Division of The
  Institute of Navigation (ION GNSS+)}.

\bibitem[Skytruth, 2019]{SkytruthJamming}
Skytruth (2019).
\newblock {Systematic GPS Manipulation Occurring at Chinese Oil Terminals and
  Government Installations}.
\newblock
  \url{https://skytruth.org/2019/12/systematic-gps-manipulation-occuring-at-chinese-oil-terminals-and-government-installations}.

\bibitem[Spanghero and Papadimitratos, 2023]{SpangheroPP:C:2023}
Spanghero, M. and Papadimitratos, P. (2023).
\newblock {Detecting GNSS misbehavior leveraging secure heterogeneous time
  sources}.
\newblock In {\em IEEE/ION Position, Location and Navigation Symposium
  (PLANS)}, Monterey, California.

\bibitem[Spens et~al., 2022]{Spensnavi.537}
Spens, N., Lee, D.-K., Nedelkov, F., and Akos, D. (2022).
\newblock Detecting gnss jamming and spoofing on android devices.
\newblock {\em NAVIGATION: Journal of the Institute of Navigation}, 69(3).

\bibitem[Spirent, 2017]{spirentSpoofing}
Spirent (2017).
\newblock {DEFCON25: GPS time spoofing now “simple party trick” -
  researcher}.
\newblock \url{https://www.spirent.com/blogs/defcon-25}.

\bibitem[Testnor, 2024a]{jammertest-event}
Testnor (2024a).
\newblock {Jammertest - The world’s largest open jamming and spoofing test}.
\newblock \url{https://jammertest.no/}.

\bibitem[Testnor, 2024b]{jammertest_test_catalog}
Testnor (2024b).
\newblock {Jammertest Transmission plan}.
\newblock
  \url{https://github.com/NPRA/jammertest-plan/blob/main/Testcatalog.pdf?ref=jammertest.no}.

\bibitem[ublox, 2022]{ubloxf9p}
ublox (2022).
\newblock {ZED-F9P module Product Datasheet - U-Blox}.
\newblock
  \url{https://www.u-blox.com/sites/default/files/ZED-F9P-04B_DataSheet_UBX-21044850.pdf}.

\bibitem[ublox, 2024]{ubx_osnma}
ublox (2024).
\newblock Ublox - osnma.
\newblock \url{https://www.u-blox.com/en/technologies/osnma-galileo-spoofing}.

\bibitem[Yan et~al., 2019]{8915828}
Yan, W., Bastos, L., and Magalhães, A. (2019).
\newblock Performance assessment of the android smartphone’s imu in a
  gnss/ins coupled navigation model.
\newblock {\em IEEE Access}, 7:171073--171083.

\bibitem[Zhang et~al., 2022]{ZhangLP:J:2022}
Zhang, K., Larsson, E.~G., and Papadimitratos, P. (2022).
\newblock {Protecting GNSS Open Service Navigation Message Authentication
  Against Distance-Decreasing Attacks}.
\newblock {\em IEEE Transactions on Aerospace and Electronic Systems (IEEE
  TAES)}, 58(2):1224--1240.

\end{thebibliography}
